\documentclass[amsmath,amssymb,pra,twocolumn,showpacs,nofootinbib,letter]{revtex4-1}

\usepackage{graphicx}
\usepackage[colorlinks=true,linkcolor=blue,citecolor=blue,urlcolor=blue]{hyperref}
\usepackage{enumerate}

\newcommand{\om}{\omega}
\newcommand{\Om}{\Omega}
\newcommand{\ii}{i}
\newcommand{\ee}{e}
\newcommand{\dd}{d}
\newcommand{\dx}{\dd x}

\newcommand{\ommin}{\om_{\rm min}}
\newcommand{\ommax}{\om_{\rm max}}

\newcommand{\enquote}{}

\begin{document}

\title{Quantum vacuum emission in a nonlinear optical medium illuminated by a strong laser pulse}

\author{Stefano Finazzi}
\email{finazzi@science.unitn.it}
\author{Iacopo Carusotto}
\email{carusott@science.unitn.it}
\affiliation{INO-CNR BEC Center and Dipartimento di Fisica, Universit\`a di Trento, via Sommarive 14, 38123 Povo-Trento, Italy}

\begin{abstract}
A strong light pulse propagating in a nonlinear Kerr medium produces a change in the refractive index, which makes light travel at different speeds inside and outside the pulse. By tuning the pulse velocity, an analog black hole horizon can be obtained in a suitable frequency window.
In this paper, we develop a quantum theory of light propagation for this system, including the frequency dispersion of the refractive index of the medium by coupling the electromagnetic field to matter polarization fields.
In a configuration with a single black hole horizon, the spectrum of spontaneously emitted particles presents some similarities with Hawking radiation. Furthermore, even in horizonless systems spontaneous vacuum emission is still possible due to the dispersive nature of the medium, yet with dramatically different spectral properties.
\end{abstract}

\date{\today}
\maketitle

\section{Introduction}

Hawking radiation~\cite{hawkingnat,hawking75} is the quantum production of particles from vacuum fluctuations due to the presence of a black hole horizon in a curved stationary geometry. 
In the pioneering work~\cite{unruh}, it was shown that this prediction does not reside on the peculiar dynamical features of a general-relativistic space-time, but only on the kinematic properties of a quantum field living in a curved space-time.
As a result, analogous quantum vacuum emission phenomena have been anticipated to occur in several physical systems, ranging from flowing fluids and superfluids, to ion rings, to nonlinear optical systems~\cite{lr}.
In particular, the advanced techniques of pulse manipulation and light detection that have been developed in the recent years put nonlinear optical systems among the most promising candidates for the realization of analog models of gravitational systems. 

In addition to the many proposals that appeared in the last few years to observe analog Hawking radiation in optical systems~\cite{ulf_science,transfoptics,dario,fleurovbarad}, the first claim of observation of an analog Hawking radiation in a laboratory was reported in Refs.~\cite{faccioexp,faccionjp} using the refractive index change induced by a strong laser pulse propagating across a nonlinear dielectric medium.
The velocity of the pulse can be tuned either by changing the laser wavelength or by using an axicon with different angles to produce a Bessel beam~\cite{durninprl,gori,mcdonald}). In this way the speed of optical photons inside the pulse can be made smaller than the velocity of the pulse itself, while the speed of photons outside the pulse remains larger. Consequently, the boundaries of the pulse appear as analog horizons, as seen from the frame comoving with the pulse. Based on the analogy with gravity, they are therefore expected to emit analog Hawking radiation.
Unfortunately, the experimental observation in Refs.~\cite{faccioexp,faccionjp} of Hawking emission in this system is still considered as controversial by some authors, who recently raised a few issues~\cite{comment,reply}. In spite of attempts at alternative theoretical interpretation~\cite{angus,unruhuniverse}, no satisfactory complete model of the experimental observations has yet been found, in particular for what concerns the crucial effect of the nontrivial dispersion of the refractive index.

Different from previous studies that reside deep in the analogy with gravitational physics, the present work aims to develop a microscopical quantum optical model of light propagation in a nonlinear dielectric modulated by the passage of a high-intensity laser pulse. 
The structure of radiative electromagnetic modes is described by using the Lagrangian of the electromagnetic field coupled with three polarization fields. In this way, we are able to canonically quantize a theory which exactly reproduces the complete Sellmeier dispersion of transparent materials such as the fused silica used in the experiment~\cite{faccioexp}. This allows us to describe the system in a far more realistic way than existing works based on simplified subluminal dispersions~\cite{Scott_thesis,robertson} and obtain quantitative predictions for the spectrum of spontaneously emitted photons in a one-dimensional geometry, under the only simplifying assumptions that the strong pulse propagates through the medium in a steady and rigid way and that the jump in the spatial profile of the refractive index is very sharp.

One of the most important conclusions of our work is that the highly nontrivial dispersion relation makes the analogy with standard quantum field theory in curved space-time much weaker than in other analog systems considered in the literature, such as Bose-Einstein condensates~\cite{lr}: For instance, in the present nonlinear optical context, an analog horizon can be defined only in a finite range of frequencies~\cite{kinetics}  that does not extend to the long-wavelength limit where the analog geometry is generally defined. Nonetheless, the properties of the quantum vacuum emission in this frequency range still share several features of standard Hawking radiation.

Furthermore, in contrast to nondispersive media, where a necessary and sufficient condition to trigger emission processes {\it \`a la} Hawking is the presence of analog horizons, quantum vacuum radiation occurs in dispersive media even in the absence of any horizon as soon as the dispersion allows for modes with a negative norm but a positive frequency, as seen from the pulse comoving frame~\cite{kinetics}. With respect to the above-mentioned Hawking-like emission channel, this additional emission is however generally much weaker.

The structure of the article is as follows. In Sec.~\ref{sec:settings}, we start from the Lagrangian to derive and solve the equation of motion for the electromagnetic field coupled to matter polarization fields. In Sec.~\ref{sec:modes}, the properties of the eigenmodes are investigated in the most significant configuration with a single analog black hole horizon, and the two bases of in- and outgoing scattering modes are built. In Sec.~\ref{sec:emission}, the spectrum of emitted particles is computed in the case of a large refractive index jump between the two sides of the analog black hole horizon. The more realistic case of a small refractive index jump is studied in Sec.~\ref{sec:otherconf} along with the horizonless configuration recently realized in experiment~\cite{faccioexp}. Conclusions are finally drawn in Sec.~\ref{sec:conclu}.

\section{The general framework}
\label{sec:settings}

In this section we develop the quantum theory used to describe light propagation in a dielectric medium with a refractive index modulation that moves in space at a uniform velocity $v$: In the experiments, the refractive index change is generated by the strong pulse via the Kerr nonlinearity of the medium. At the simplest level of approximation, one can assume that the refractive index change follows locally in space and instantaneously in time the intensity profile of the pump pulse~\cite{butchercotter}.
 
The optical properties of the medium are described in the spirit of the Hopfield model~\cite
{hopfield} in terms of an electromagnetic field interacting with a polarization field. To closely reproduce the Sellmeier dispersion relation of typical transparent dielectrics such as the fused silica used in the experiment of Ref.~\cite{faccioexp}, the polarization of the medium has to show three poles with different strengths and frequencies. The effect of the passage of the pulse in the medium is then modeled as a spatiotemporal variation of these quantities, considered as external parameters. More sophisticated models where the internal level structure of the emitters is explicitly taken into account were developed in Refs.~\cite{iacopothree,subband,iacopo_PRA_backreaction} but the complexity of such a description goes far beyond the scope of the present work.

For the sake of simplicity, the strong pulse is assumed to propagate across the medium at a constant speed and to maintain a constant shape in space.
This approximation prevents us from describing possible time-dependent effects, as studied in Ref.~\cite{angus}.
We also restrict our description to a one-dimensional geometry where light modes propagate parallel to the direction of the pulse velocity. The refractive index jump is assumed to be localized at the sharp interface between two asymptotic homogeneous regions. More general configurations with smooth refractive index profiles will be the subject of future work.

\subsection{The field equation}
\label{subsec:field}

In the laboratory reference frame, the Lagrangian density in one dimension of the electromagnetic field coupled to $N$ polarization fields $P_i$ is
\begin{multline}\label{eq:lagrangian}
 {\cal L}_l=\frac{{(\partial_{T} A)}^2}{8\pi c^2}-\frac{{(\partial_{X} A)}^2}{8\pi}\\
 +\sum_{i=1}^N\left(\frac{{(\partial_{T} P_i)}^2}{2\beta_i\Om_i^2}-\frac{P_i^2}{2\beta_i}+\frac{1}{c}A\,{\partial_{T} P_i}\right),
\end{multline}
where $A$ and $P_i$ oscillate in a direction orthogonal to their propagation direction.
In this simple model the polarization is described by fields of harmonic oscillators, with elastic constant $\beta^{-1}$ and inertia $(\beta_i\Om_i^2)^{-1}$. For the sake of simplicity, in this paper we restrict to the case $N=3$, which suitably describes the dispersion relation in fused silica, the material used in the experiment of Ref.~\cite{faccioexp}.

A propagating strong light pulse causes a local perturbations of the parameters $\beta_i$ and $\Om_i$.
Consequently, the system is stationary when observed in the reference frame comoving with the pulse at velocity $v$.
Thus, it is convenient to transform the laboratory coordinates $X$ and $T$ to the comoving coordinates $x$ and $t$, by applying a Lorentz boost $\Lambda$ with velocity $v$
\begin{equation}\label{eq:boost}
  t = \gamma[T-vX/c^2],\quad
  x = \gamma[X-vT],
\end{equation}
where $\gamma=1/\sqrt{1-v^2/c^2}$. Coherently, differential operators transform as
\begin{equation}
  \partial_{T} = \gamma[\partial_{t}-v\partial_{x}],\quad
  \partial_{X} = \gamma[\partial_{x}-v\partial_{t}/c^2].
\end{equation}
Furthermore, we treat $A$ and $P_i$ as scalar fields; that is, we do not transform them under the boost. Although this might appear not correct, it is completely legitimate, as shown in Appendix~\ref{app:scalars}.

The transformed Lagrangian density is
\begin{multline}\label{eq:lagrangianboost}
 {\cal L}=\frac{{\dot A}^2}{8\pi c^2}-\frac{{A'}^2}{8\pi}
 +\sum_{i=1}^3\left[
 \frac{\gamma^2}{2\beta_i\Om_i^2}({\dot P}_i-vP_i')^2\right.\\\left.
-\frac{P_i^2}{2\beta_i}+\frac{\gamma}{c}A ({\dot P_i}-vP_i')^2\right],
\end{multline}
where dot and prime denote derivation with respect to $t$ and $x$, respectively.

As usual, the conjugate momenta are obtained by varying the Lagrangian
\begin{equation}
 L=\int \dx\,{\cal L}
\end{equation}
with respect to the time derivatives of $A$ and $P_i$:
\begin{equation}\label{eq:momenta}
 \Pi_A=\frac{\dot A}{4\pi c^2},\quad\Pi_{P_i}=\frac{\gamma^2}{\beta_i\Om_i^2}({\dot P}_i-vP_i')+\frac{\gamma}{c}A.
\end{equation}
We can impose canonical commutation relations on $A$ and $P_i$ (see Appendix~\ref{app:scalars}),
\begin{equation}\label{eq:commutators}
\begin{aligned}
 &[A(x),\Pi_A(x')]=\ii\hbar\,\delta(x-x'),\\
 &[P_i(x),\Pi_{P_j}(x')]=\ii\hbar\,\delta_{ij}\delta(x-x'),
\end{aligned}
\end{equation}
and all the other commutators vanish.

The Hamiltonian density
\begin{equation}
 {\cal H}=\frac{1}{2}\left[\dot A\, \Pi_A+\Pi_A\dot A+\sum_{i=1}^3\left(\dot P_i\,\Pi_{P_i}+\Pi_{P_i}\dot P_i\right)\right]-{\cal L}
\end{equation}
becomes
\begin{multline}
 {\cal H}=2\pi c^2\,\Pi_A^2+\frac{{A'}^2}{8\pi}+\sum_{i=1}^3\left[\frac{\beta_i\Om_i^2}{2\gamma^2}\left(\Pi_{P_i}-\frac{\gamma}{c}A\right)^2
 \right.\\\left.
 +\frac{P_i^2}{2\beta_i}+\frac{1}{2}(P'_i\,\Pi_{P_i}+\Pi_{P_i}P'_i)\right],
\end{multline}
and the Hamilton equations are derived by the commutators of the fields and their conjugate momenta with the Hamiltonian
\begin{equation}
 H=\int \dx\,{\cal H}.
\end{equation}
We obtain
\begin{align}
 &\dot A=4\pi c^2\,\Pi_A,\label{eq:system1}\\
 &\dot P_i=\frac{\beta_i\Om_i^2}{\gamma^2}\left(\Pi_{P_i}-\frac{\gamma}{c}A\right)+vP_i',\label{eq:system2}\\
 &{\dot \Pi}_A=\frac{A''}{4\pi}+\sum_{i=1}^3\left[\frac{\beta_i\Om_i^2}{\gamma c}\left(\Pi_{P_i}-\frac{\gamma}{c}A\right)\right],\label{eq:system3}\\
 &{\dot \Pi}_{P_i}=-\frac{P_i}{\beta_i}+\partial_{x}\left(v\Pi_{P_i}\right).\label{eq:system4}
\end{align}
Note that one might have obtained the first two equations directly from the definition of conjugate momenta~\eqref{eq:momenta}.

It is now convenient to define the eight-dimensional vector
\begin{equation}
 V=
 \begin{pmatrix}
  A & P_1 & P_2 & P_3 & \Pi_A & \Pi_{P_1} & \Pi_{P_2} & \Pi_{P_3}
 \end{pmatrix}^T
\end{equation}
and the matrix
\begin{equation}
 \eta=
 \begin{pmatrix}
  0&I_4\\-I_4&0
 \end{pmatrix},
\end{equation}
where $I_4$ is the $4\times4$ identity matrix. With this notation the Hamilton equations can be written in a compact form as
\begin{equation}\label{eq:compact}
 \dot V=\eta(\nabla_V {\cal H}).
\end{equation}

Moreover, a scalar product
\begin{equation}\label{eq:scalar}
 \langle V_1,V_2\rangle=\frac{\ii}{\hbar}\int \dx\, V_1^\dagger(x,t)\,\eta\, V_2(x,t)
\end{equation}
can be defined on the set of the solutions of Eq.~\eqref{eq:compact}, generalized to complex values.
By virtue of Eq.~\eqref{eq:compact}, using $\eta^2=-I_8$, $\eta^\dagger\eta=I_8$ and the fact that the Hamiltonian density ${\cal H}$ is quadratic in the fields and their momenta, this scalar product is conserved by time evolution:

\begin{multline}
 \partial_{t}\langle V_1,V_2\rangle
 =\frac{\ii}{\hbar}\int \dx\left[ \partial_{t}(V_1^\dagger)\,\eta\, V_2+V_1^\dagger\,\eta\, \partial_{t}(V_2)\right]\\
 =\frac{\ii}{\hbar}\int \dx\left[ \left.(\nabla_V{\cal H})^\dagger\right|_{(V=V_1)} V_2\right.\quad\\\left.
 -V_1^\dagger \left.(\nabla_V{\cal H})^\dagger\right|_{(V=V_2)}\right]=0.
\end{multline}

Being the system stationary in the reference system comoving with the pulse, it is convenient to expand the real field $V$ on a basis of frequency eigenmodes $V_\omega$, rather than, as usually done, on a basis of wave-vector eigenmodes:
\begin{equation}\label{eq:expansion}
 V=\int \dd \om\sum_\alpha \left(V_{\om}^{\alpha}{\hat a}_{\om}^\alpha+V_{\om}^{\alpha*}{\hat a}_{\om}^{\alpha\dagger}\right),
\end{equation}
where
\begin{equation}\label{eq:a}
 \hat a_\om^\alpha=\langle V_\om^\alpha,V\rangle,
\end{equation}
the label $\alpha$ denotes various modes with the same eigenfrequency $\om$, $V_{\om}^{\alpha}$ are properly normalized (see Appendix~\ref{app:norm}) with respect to the norm induced by the scalar product defined in Eq.~\eqref{eq:scalar}, and the integral generally includes both positive- and negative-frequency modes with positive norm.

\subsection{Homogeneous systems}
\label{subsec:homogeneous}

In the asymptotic regions, far from the perturbation, the system is homogeneous and the parameters $\Om_i$, $\beta_i$, and $v$ are constant in both time and space. In this situation one can chose $V_{\om}^\alpha$ as momentum eigenmodes,
\begin{equation}\label{eq:momentumeigenmode}
 V_{\om}^{\alpha}(x,t)=\ee^{-\ii\om t+\ii k_\alpha x} \bar V_{\om}^{\alpha},
\end{equation}
where $\bar V_{\om}^{\alpha}$ is a vector of constant $\mathbb{C}$ numbers, satisfying
\begin{equation}\label{eq:fouriersystem}
 -\ii\om\bar V_{\om}^{\alpha}=\eta\, {\cal K}(k_\alpha)\,\bar V_{\om}^{\alpha},
\end{equation}
and
\begin{widetext}
\begin{equation}
 {\cal K}(k_\alpha)=
 \begin{pmatrix}
  k_\alpha^2/4\pi+\sum_{i=1}^3\beta_i\Om_i^2/c^2  & 0 & 0 & 0 & 0 & -\beta_1\Om_1^2/\gamma c & -\beta_2 \Om_2^2/\gamma c & -\beta_3\Om_3^2/\gamma c\\
  0 & 1/\beta_1 & 0 & 0 & 0 & -\ii k_\alpha v & 0 & 0 \\
  0 & 0 & 1/\beta_2 & 0 & 0 & 0 & -\ii k_\alpha v & 0 \\
  0 & 0 & 0 & 1/\beta_3 & 0 & 0 & 0 & -\ii k_\alpha v \\
  0 & 0 & 0 & 0 & 4\pi c^2 & 0 & 0 & 0 \\
  -\beta_1\Om_1^2/\gamma c & + \ii k_\alpha v & 0 & 0 & 0 & \beta_1\Om_1^2/\gamma^2 & 0 & 0 \\
  -\beta_2\Om_2^2/\gamma c & 0 & + \ii k_\alpha v & 0 & 0 & 0 & \beta_2\Om_2^2/\gamma^2 & 0 \\
  -\beta_3\Om_3^2/\gamma c & 0 & 0 & + \ii k_\alpha v & 0 & 0 & 0 & \beta_3\Om_3^2/\gamma^2\\
 \end{pmatrix}
\end{equation}
\end{widetext}
Equation~\eqref{eq:fouriersystem} can be written as
\begin{equation}
 (\ii\eta\, {\cal K}(k_\alpha)-\om I_8)\bar V_{\om}^{\alpha}=0,
\end{equation}
which has nonvanishing solutions if and only if
\begin{equation}
 \mbox{det}[\ii\eta\, {\cal K}(k_\alpha)-\om I_8]=0.
\end{equation}
The computation of the above determinant yields the dispersion relation
\begin{equation}\label{eq:dispersion}
 c^2 k_\alpha^2 = \om^2 +\sum_{i=1}^3\frac{4\pi\beta_i\, \gamma^2(\om+v k)^2}{1-\gamma^2(\om+v k)^2/\Om_i^2},
\end{equation}
that has, in general, eight solutions $k_\alpha$ for each value of the frequency $\omega$.

Applying the boost of Eq.~\eqref{eq:boost} to $\om$ and $k$, the corresponding frequency $\Om$ and wave number $K$, measured in the laboratory reference frame, are
\begin{equation}\label{eq:boosfrequency}
  \Om = \gamma[\om +v k],\quad
  K = \gamma[k+v\om/c^2].
\end{equation}
Equation~\eqref{eq:dispersion} becomes
\begin{equation}\label{eq:sellmeier}
  c^2 K^2 = \Om^2\left[1 +\sum_{i=1}^3\frac{4\pi\beta_i}{1-\Om^2/\Om_i^2}\right],
\end{equation}
which is the well-known Sellmeier dispersion relation~\cite{sellmeier,refractiveindex}.

The solutions of the system~\eqref{eq:fouriersystem} (eigenmodes of $\ii\eta\cal{K}$) are
\begin{equation}\label{eq:solution}
 \bar V_\om^\alpha = C_\om^\alpha
 \begin{pmatrix}
  c\\
  \ii\beta_1\,\gamma(\om+v k_\alpha)[1-\gamma^2(\om+v k)^2/\Om_1^2]^{-1}\\
  \ii\beta_2\,\gamma(\om+v k_\alpha)[1-\gamma^2(\om+v k)^2/\Om_2^2]^{-1}\\
  \ii\beta_3\,\gamma(\om+v k_\alpha)[1-\gamma^2(\om+v k)^2/\Om_3^2]^{-1}\\
  -\ii\om/4\pi c\\
  \gamma[1-\gamma^2(\om+v k)^2/\Om_1^2]^{-1}\\
  \gamma[1-\gamma^2(\om+v k)^2/\Om_2^2]^{-1}\\
  \gamma[1-\gamma^2(\om+v k)^2/\Om_3^2]^{-1}
 \end{pmatrix},
\end{equation}
where the constant $C_\om^\alpha$ is computed in Appendix~\ref{app:norm},
\begin{equation}
 |C_\om^\alpha|^2
 =
\hbar\left|
c^2 k_\alpha -
v\sum_{i=1}^3\frac{4\pi\beta_i\gamma^2(\om-vk_\alpha)}{[1-\gamma^2(\om-vk_\alpha)^2/\Om_i^2]^2}
 \right|^{-1},
\end{equation}
using the normalization condition
\begin{equation}\label{eq:normalization}
  |\left\langle V_{\om_1}^{\alpha_1},V_{\om_2}^{\alpha_2}\right\rangle|
 =\delta(\om_2-\om_1)\,\delta_{\alpha_2\alpha_1}.
\end{equation}
The modulus is needed since some modes have negative norm, because the scalar product of Eq.~\eqref{eq:scalar} is not positive definite. In particular, it is possible to show (see Appendix~\ref{app:norm}) that
the norm of a mode is positive when its laboratory frequency $\Om$ is positive, whereas it is negative when $\Om<0$, independently of the value of $K$.

The expansion of Eq.~\eqref{eq:expansion}, which contains positive-norm eigenmodes $V_\om^\alpha$ with either positive or negative frequency $\om$, can be rewritten by considering only positive frequencies $\om$. As a consequence, negative-norm eigenmodes must be included in the integral. Indeed, for a given positive value of the comoving frequency $\om$, the dispersion relation~\eqref{eq:dispersion} admits solutions for $k$, such that $\Om=\gamma(\om+ vk)$ is negative and the associated modes have negative norm.
This implies that, in the expansion of the field $V$, the Fock operators associated with those positive-$\om$ modes are not destruction operators but instead creation operators. Naming $P$ the set of positive-norm modes $V_{\om}^{\alpha}$, labeled by $\alpha$, and $N$ the set of negative-norm modes $V_{\om}^{\tilde\alpha}$, labeled by $\tilde\alpha$, $V$ becomes
\begin{multline}\label{eq:expansionadagger}
 V=\int_0^\infty\dd
\om\,\ee^{-\ii\om t}\left(\sum_{\alpha\in P} \ee^{+\ii k_\alpha x} \bar V_{\om}^{\alpha}{\hat a}_{\om}^\alpha
\right.\\\left.
+\sum_{\tilde\alpha\in N}\ee^{+\ii k_{\tilde\alpha} x} \bar V_{\om}^{\tilde\alpha}
{\hat a}_{\om}^{\tilde\alpha\dagger}\right)+\mbox{H.c.},
\end{multline}
where H.c. stands for Hermitian conjugate. Note that the positive-frequency part of the field (i.e., evolving with $\ee^{-\ii\om t}$) mixes creation $\hat a_\om^\alpha$ and destruction $\hat a_{\om}^{\tilde\alpha\dagger}$ operators.

Using Eq.~\eqref{eq:a} and the normalization of Eq.~\eqref{eq:normalization} for the frequency eigenmodes $V_\om^\alpha$, it is easy to check that $\hat a_\om^\beta$ and $\hat a_\om^{\beta\dagger}$ are creation and destruction operators, since their commutation relations, implied by the canonical commutators of the fields given in Eq.~\eqref{eq:commutators}, are
\begin{equation}\label{eq:fockcomm}
 \left[\hat a_\om^\beta,\hat a_{\om'}^{\beta'\dagger}\right]=\delta(\om-\om')\delta_{\beta\beta'},
\end{equation}
where now $\beta$ can indifferently belong to either $P$ or $N$.

\subsection{Matching conditions}
\label{subsec:matching}

In this paper, we consider only configurations with a single analog black hole horizon.
For the sake of simplicity we model this system with two homogeneous half-line regions (representing, respectively, the interior and the exterior of the analog black hole) connected by a discontinuity at $x=0$.
On the two sides of the discontinuity, the elastic constant $\beta_i^{-1}$ takes different constant values
\begin{equation}
 \beta_i=\beta_{i,L}\theta(-x)+\beta_{i,R}\theta(x).
\end{equation}
We also impose that the inertia $(\beta_i\Om_i^2)^{-1}$, physically corresponding to the masses of the oscillator fields [see Eq.~\eqref{eq:lagrangian}], is the same in the two regions; that is, $\Om_i$ must vary accordingly to
\begin{equation}
 \beta_{i,L}\Om_{i,L}^2=\beta_{i,R}\Om_{i,R}^2.
\end{equation}
In the context of analog systems, the steplike profile has been proved to provide reliable results when the parameters do not vary very much between the two homogeneous regions~\cite{2dplots}. This is indeed the case in the optical system considered in Ref.~\cite{faccioexp} where the relative difference in the refractive index in the two regions is of the order of $0.1\%$.

In this geometry, a frequency eigenmode $V_\om^\alpha$ [see Eq.~\eqref{eq:expansion}] can be written as
\begin{equation}\label{eq:leftright}
 V_\om^\alpha=\sum_{\alpha}L^{\alpha}_\om\, V_{\om,L}^{\alpha}\,\theta(-x)
 +\sum_{\alpha}R^{\alpha}_\om\, V_{\om,R}^{\alpha}\,\theta(x),
\end{equation}
where $L^{\alpha}_\om$ and $R^{\alpha}_\om$ are constant and $V_{\om,L}^{\alpha}$ and $V_{\om,R}^{\alpha}$ are
frequency-momentum eigenmodes, as in Eq.~\eqref{eq:momentumeigenmode}; that is, they are
solutions of the field equation Eq.~\eqref{eq:compact} in the homogeneous left ($x<0$) and right ($x>0$) regions, respectively.

The relations between $L^{\alpha}_\om$'s and $R^{\alpha}_\om$'s, are determined by solving the field equation in a neighborhood of $x=0$. Writing $V_\om^\alpha$ of Eq.~\eqref{eq:leftright} in a more compact form,
\begin{equation}\label{eq:Vcompact}
 V_\om^\alpha=V_{L}\,\theta(-x)+V_{R}\,\theta(x),
\end{equation}
its first and second spatial derivatives are
\begin{align}
 &{V_\om^{\alpha}}'=V_{L}'\,\theta(-x)+V_{R}'\,\theta(x)+(V_R-V_L)\delta(x),\\
  &
 {V_\om^{\alpha}}''=V_{L}''\,\theta(-x)+V_{R}''\,\theta(x)+2(V_R'-V_L')\delta(x)\nonumber\\
 &\qquad\qquad+(V_R-V_L)\delta'(x).
\end{align}
We now put the above expressions in Eqs.~\eqref{eq:system1}, \eqref{eq:system2}, \eqref{eq:system3}, and~\eqref{eq:system4}, and group all the terms with $\theta(x)$, $\theta(-x)$, $\delta(x)$, and $\delta'(x)$.
The terms with $\theta(x)$ and $\theta(-x)$ trivially give two sets of equations, separately satisfied by $V_L$ and $V_R$ in the homogeneous left and right regions.
These equations are solved as in Sec.~\ref{subsec:homogeneous}.
For instance, Eq.~\eqref{eq:system1} gives
\begin{equation}\label{eq:firstleftright}
 \dot A_R=4\pi c^2\,\Pi_{A,R},\quad \dot A_L=4\pi c^2\,\Pi_{A,R}.
\end{equation}

The connection formulas between the two asymptotic regions are given instead by $\delta(x)$ and $\delta'(x)$ terms. Equation~\eqref{eq:system1} does not contain any of them. Equation~\eqref{eq:system2} has a term with $\delta(x)$
\begin{equation}
 v\left(P_{i,R}-P_{i,L}\right)\delta(x)=0,
\end{equation}
yielding
\begin{equation}
 v\left[P_{i,R}(0)-P_{i,L}(0)\right]=0,
\end{equation}
which has two solutions:
\begin{equation}\label{eq:solP}
 v=0\quad\mbox{or}\quad P_{i,R}(0)=P_{i,L}(0).
\end{equation}
Thus, when $v\neq0$, $P_i$'s must be continuous at $x=0$.

Equation~\eqref{eq:system3} contains a second derivative of $A$, generating both a term with $\delta(x)$ and one with $\delta'(x)$. The associated equations are
\begin{equation}
 A_R(0)=A_L(0),\quad A_R'(0)=A_L'(0),
\end{equation}
so that both $A$ and $A'$ are continuous in $x=0$; that is, the magnetic field $B$ is continuous.
Consequently, because of Eq.~\eqref{eq:firstleftright}, also $\Pi_A$ and $\Pi_A'$ are continuous in $x=0$. Since the electric field is proportional to the time derivative of $A$, this implies that both $E$ and its first spatial derivative are continuous in $x=0$.

Finally, Eq.~\eqref{eq:system4} contains the first derivative of $\Pi_P$ ($v$ is constant), generating a $\delta(x)$ term that yields
\begin{equation}\label{eq:solPiP}
 v=0\quad\mbox{or}\quad \Pi_{P_i,L}(0)=\Pi_{P_i,R}(0),
\end{equation}
and, consequently, $\Pi_{P_i}$'s are continuous at $x=0$ when $v\neq0$.

Summarizing, we proved that
\begin{enumerate}[(i)]
 \item $A$, $\Pi_A$, and their first and second derivatives are continuous in $x=0$,
 \item if $v\neq0$, $P_i$ and $\Pi_{P_i}$ are continuous.
\end{enumerate}
To completely fix the relations between modes in the left and right regions, one must determine the matching conditions of $P_i'$ and $\Pi_{P_i}'$, and also those of of $P_i$ and $\Pi_{P_i}$ if $v=0$.

\setcounter{paragraph}{0}
\paragraph{$v\neq 0$.}
The $\theta(-x)$ and $\theta(x)$ terms of Eq.~\eqref{eq:system2} are
\begin{equation}\label{eq:secondleftright}
\begin{aligned}
 vP_{i,L}'&=\dot P_{i,L}-\frac{\beta_{i,L}\Om_{i,L}^2}{\gamma^2}\left(\Pi_{P_i,L}-\frac{\gamma}{c}A_L\right),\\
 vP_{i,R}'&=\dot P_{i,R}-\frac{\beta_{i,R}\Om_{i,R}^2}{\gamma^2}\left(\Pi_{P_i,R}-\frac{\gamma}{c}A_R\right).
\end{aligned}
\end{equation}
Since $A$, $\Pi_{A}$, and $\dot P_i=\ii\om P_i$ are continuous, and $\beta_i\Om_i^2$ are equal on both sides of the discontinuity, Eq.~\eqref{eq:secondleftright} implies that also $P_{i}'$ is continuous at $x=0$.

Similartly, the matching condition for $\Pi_{P_i}'$ is found from the $\theta(-x)$ and $\theta(x)$ terms of Eq.~\eqref{eq:system4}:
\begin{equation}\label{eq:fourthleftright}
\begin{aligned}
  \beta_{i,L}\left({\dot \Pi}_{P_{i},L}-v\Pi_{P_i,L}'\right)=-P_{i,L},\\
  \beta_{i,R}\left({\dot \Pi}_{P_{i},R}-v\Pi_{P_i,R}'\right)=-P_{i,R}.
\end{aligned}
\end{equation}
By continuity of $P_i$,
\begin{equation}\label{eq:PiP'}
 \beta_{i}\left({\dot \Pi}_{P_{i}}-v\Pi_{P_i}'\right)=-P_{i}
\end{equation}
is continuous at $x=0$ and, accordingly, $\Pi_{P_i}'$ is discontinuous.

The continuity of $\dot P_i$ and $P_i'$ implies that the derivative of $P_i$ with respect to the laboratory time $T$,
\begin{equation}
 \partial_T P_i=\gamma(\dot P_i-v P_i'),
\end{equation}
is continuous. Analogously, Eq.~\eqref{eq:PiP'} implies that
\begin{equation}
 \beta_i\partial_T \Pi_{P_i}
\end{equation}
is continuous at $x=0$. The physical meaning of this condition is evident when the second equation of~\eqref{eq:momenta} is rewritten as
\begin{equation}
 \Pi_{P_i}=\frac{\gamma}{\beta_i\Om_i^2}\partial_T P_i+\frac{\gamma}{c}A.
\end{equation}
Differentiating with respect to $T$,
\begin{equation}
 \partial_T\Pi_{P_i}=\frac{\gamma}{\beta_i\Om_i^2}\partial_T^2 P_i+\frac{\gamma}{c}\partial_T A,
\end{equation}
where we used the constancy of $\beta_i\Om_i^2$.
Multiplying both sides of this equation by $\beta_i$, using Eq.~\eqref{eq:PiP'} in the form
\begin{equation}
 \beta_{i}\,\partial_T\Pi_{P_i}=-\gamma P_{i},
\end{equation}
and noting that $E=-\partial_T A/c$ is the electric field, one obtains
\begin{equation}\label{eq:oscillator}
 \partial_T^2 P_i= -\Om_i^2 P_i +\beta_i\Om_i^2 E.
\end{equation}
This equation describes the dynamics of a harmonic oscillator of frequency $\Omega_i$ and mass $(\beta_i\Omega_i^2)^{-1}$, which is forced by $E$. When the perturbation, caused by the laser pulse, passes at some $X=X_0$, at time $T=T_0=X_0/v$, the mass $(\beta_i\Omega_i^2)^{-1}$ remains unchanged, but the oscillator frequency $\Om_i$ changes from $\Om_{i,R}$ to $\Om_{i,L}$. As a consequence of Eq.~\eqref{eq:oscillator}, $P_i$ and $\partial_T P_i$ are unchanged as the perturbation arrives at $X=X_0$, but $\partial_T^2 P_i$ has a finite jump; that is, it is discontinuous at $X=X_0$ and $T=T_0$.

\paragraph{$v=0$.}
In this case, Eqs.~\eqref{eq:system2} and~\eqref{eq:system4} simplify to
\begin{equation}\label{eq:v0}
 \dot P_i={\beta_i\Om_i^2}\left(\Pi_{P_i}-\frac{A}{c}\right),\quad
 {\dot \Pi}_{P_i}=-\frac{P_i}{\beta_i}.
\end{equation}
Differentiating the first equation with respect to time, using the second equation and $E=-\dot A/c$,
\begin{equation}
 \ddot P_i=-\Om_i^2\,P_i+ {\beta_i\Om_i^2}\,E,
\end{equation}
which coincides with Eq.~\eqref{eq:oscillator} because $T=t$ when $v=0$.
Since $\beta_i\Om_i^2 E$ is continuous at $x=0$,
\begin{equation}
 \ddot P_i+\Om_i^2\,P_i
\end{equation}
is continuous and, for an eigenmode of frequency $\om$,
\begin{equation}
 \left(\om^2 -\Om_{i,L}^2\right)P_{i,L}=\left(\om^2 -\Om_{i,R}^2\right)P_{i,R}
\end{equation}
at $x=0$. Thus, $P_i$ is not continuous at $x=0$, because $\Om_i$ is discontinuous.

Similarly, by deriving the second equation of~\eqref{eq:v0} with respect to time, one obtains
\begin{equation}
 {\ddot \Pi}_{P_i}=-\Om_i^2\left(\Pi_{P_i}-\frac{A}{c}\right),
\end{equation}
which implies that
\begin{equation}
 \frac{{\ddot \Pi}_{P_i}}{\Om_i^2}+\Pi_{P_i}
\end{equation}
is continuous; that is
\begin{equation}
 \left(1-\frac{\om^2}{\Om_{i,L}^2}\right)\Pi_{P_{i,L}}=\left(1-\frac{\om^2}{\Om_{i,R}^2}\right)\Pi_{P_{i,R}}
\end{equation}
at $x=0$, and $\Pi_{P_i}$ is discontinuous.

To conclude this section, it is worth checking how many constants are fixed by the matching conditions.
In Eq.~\eqref{eq:leftright}, there are 16 unknown constants $L_\om^{\alpha}$ and $R_\om^{\alpha}$, corresponding, respectively, to the 8 solutions of the dispersion relation~\eqref{eq:dispersion} in each of the two homogeneous asymptotic regions.
Moreover, for every frequency $\om$, there are, in general, 8 global solutions $V_\om^\alpha$.
Thus, those constants should be constrained by 8 equations, but we found 16 matching conditions for the continuity of $A$, $P_i$, $\Pi_A$, $\Pi_{P_i}$, and of their first derivatives. The system is apparently overdetermined. However, it is possible to show that 8 constraints are redundant.
In fact, the 8 continuity relations of  the conjugate momenta are directly implied by the 8 relations of the fields [see, for instance, Eq.~\eqref{eq:firstleftright}].

At the end of the day, for each frequency $\om$, there are 16 parameters and 8 independent constraints, leaving 8 free parameters, associated with the 8 globally defined solutions of the field equation~\eqref{eq:compact}.

\section{Mode analysis}
\label{sec:modes}

In this section we apply the quantum field formalism introduced in the previous section to build the global eigenmodes of the system by imposing the suitable matching conditions at the transition interface. The identification of the different ingoing and outgoing channels and then the calculation of the $S$ matrix describing the scattering of light at the interface are the basic ingredients to calculate the intensity and the spectrum of the quantum vacuum emission in the next section.

\subsection{Asymptotic modes}
\label{subsec:asymptotic}

%
\begin{figure*}
\centering
\includegraphics{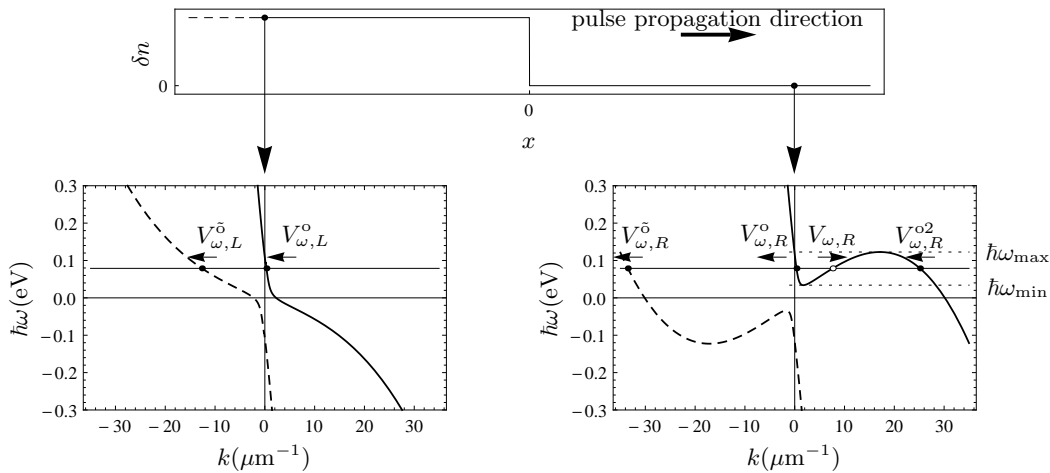}
\caption{{\it Black-hole-like configuration.} Simplified analysis of the Sellmeier dispersion relation~\eqref{eq:dispersion} in fused silica, as seen from the comoving frame. Only the optical branch is shown.
Positive (negative) laboratory frequency branches are represented by solid (dashed) curves.
The dispersion is plotted on the left ($\delta n=0.1$, left panel, interior of the analog black hole) and on the right ($\delta n =0$, right panel, exterior of the analog black hole) of a perturbation moving rightward with $v=0.66c$ in the laboratory frame (see top panel).
The black horizontal line represents a generic frequency for which there are four real solutions in the right region (right panel) and two real solutions in the left one (left panel) and the system shows an analog black hole horizon.
In the right panel, the dashed horizontal lines indicate the maximum and the minimum values of the frequency for which this behavior occurs.
Modes are labeled with the notation introduced in Sec.~\ref{sec:settings} and their propagation direction is indicated by arrows.}
\label{fig:dispsimp}
\end{figure*}

A stationary system (in the reference frame comoving with the laser pulse) made of two asymptotic homogeneous regions, connected by a transition region was first investigated under a purely kinematic perspective in Ref.~\cite{kinetics}. In particular, it was shown that it is possible to tune the velocity of propagation of the pulse in such a way that the transition between the two asymptotic flat regions looks like an analog black hole horizon. 

In that analysis, only the optical branch (corresponding to optical frequencies) of the dispersion relation was considered: Within this approximation, the dispersion relation admits four solutions of $k$ sharing the same value of the comoving frequencies $\om$. Before studying the full problem, taking into account all the branches of the dispersion relation, it is worth summarizing the most relevant aspects of the analysis in Ref.~\cite{kinetics}.

Assuming that the perturbation is moving in the positive $x$ direction ($v>0$), a frequency-dependent horizon is present for a giving comoving frequency $\om$ when, on the left of the perturbation, there are only negative group velocity (measured in the comoving frame) modes, while on the right there are both negative and positive group velocity modes.
In this case, light can propagate only leftward in the left region, both leftward and rightward in the right region. In analogy with black hole physics, the left and right regions correspond, respectively, to the interior and the exterior of a black hole, and, in the transition region, there is one point corresponding to a black hole horizon.

This physics is illustrated in Fig.~\ref{fig:dispsimp}, where the optical branch of the dispersion relation~\eqref{eq:dispersion} is plotted in the comoving frame, in both the left (left panel) and the right (right panel) regions, for a steplike pulse (top panel) moving rightward at $v=0.66c$. 
In the right region the dispersion relation is given directly by Eq.~\eqref{eq:dispersion}, while in the left region (representing the interior of a propagating pulse), the effective refractive index of optical-frequency modes has been increased by $\delta n=0.1$, by perturbing the parameters $\beta_i$ and $\Omega_i$ in the dispersion relation.
Solid (dashed) curves denote branches with positive (negative) laboratory frequency. As demonstrated in Sec.~\ref{subsec:homogeneous}, they correspond to positive- (negative-) norm modes.

The dispersion relation is solved for a given comoving frequency $\om$.
The arrows indicate the direction of propagation of the corresponding modes.
Modes are named using the notation introduced in Sec.~\ref{sec:settings}. The superscript $o$ stands for positive-norm optical branch and $\tilde o$ stands for negative-norm optical branch. The subscripts $L$ and $R$ denote, respectively, modes defined in the left and right regions.
For outer points (right panel, $\delta n=0$), the dispersion relation has four solutions. Three of them correspond to modes ($V_{\omega,R}^{\rm\tilde o}$, $V_{\omega,R}^{\rm o}$, $V_{\omega,R}^{\rm o2}$) propagating leftward from $x=+\infty$ toward the horizon.
The fourth one, denoted by an open dot, corresponds instead to an outgoing mode ($V_{\omega,R}$), with positive group velocity.
For inner points (left panel, $\delta n>0.1$) the dispersion relation has only two real solutions, both corresponding to leftgoing modes ($V_{\omega,L}^{\rm\tilde o}$, $V_{\omega,L}^{\rm o}$) which propagate from the horizon toward $x=-\infty$.
Note that the two extra solutions present only in the right region for $\ommin<\om<\ommax$ correspond, respectively, to a leftgoing ($V_{\omega,R}^{\rm o2}$) and a rightgoing ($V_{\omega,R}$) mode. 

As a result, according to the above given definition, within the frequency range $\ommin<\om<\ommax$ (where $\hbar\ommin$ and $\hbar\ommax$ are represented in the right panel with dotted lines), the discontinuity in the pulse profile represents an analog black hole horizon. Note that the dispersion relation has one solution corresponding to a negative-norm mode $V_{\omega,R}^{\rm\tilde o}$, leftward propagating from $+\infty$ to the horizon, and one solution (right panel, open dot), corresponding to a positive-norm mode $V_{\omega,R}$ propagating from the horizon to infinity in the exterior part of the analog black hole. This mode structure might originate some phenomenon similar to Hawking radiation.

In the present paper, this analysis is extended to the full Sellmeier dispersion relation, which, in general, admits eight solutions, corresponding to eight (propagating if $k$ is real) asymptotic modes (AMs) in each region ($x<0$ or $x>0$).
Proceeding in analogy with Ref.~\cite{kinetics}, one must first identify a range of frequencies $\ommin<\om<\ommax$, for which the optical branch possesses four real solutions in the right region and only two in the left one.
Second, on each side one must localize four additional solutions on the other branches of the dispersion relation and describe the associated modes.

\begin{figure*}
\begin{center}
\includegraphics[width=0.95\textwidth]{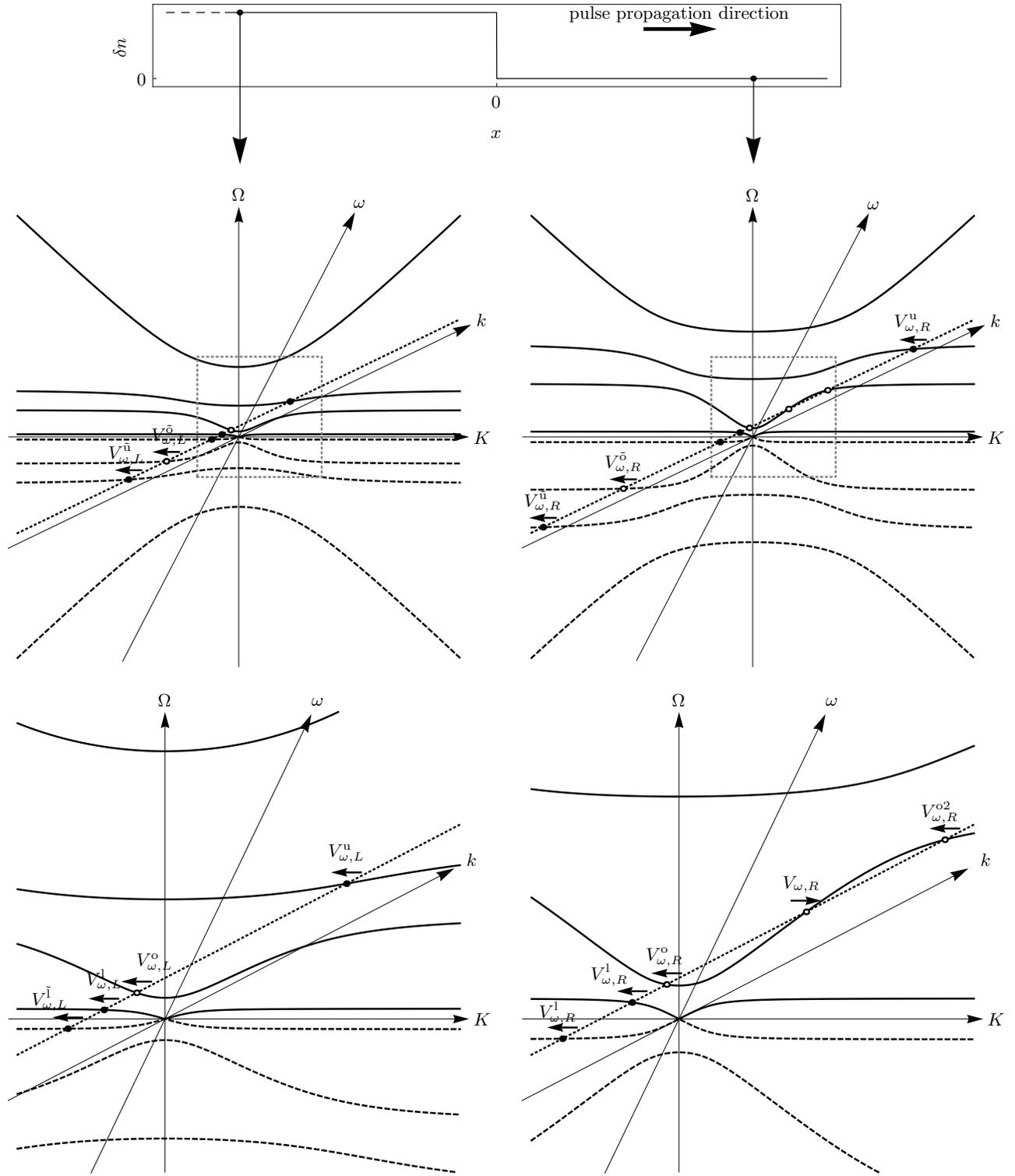}
\end{center}
\caption{{\it Black-hole-like configuration.} Graphical representation of the Sellmeier dispersion relation, as seen from the laboratory reference frame $(\Om,K)$, for $x<0$ (left panels) and $x>0$ (right panels).
The $(\om,k)$ axes of the comoving reference frame are obtained through a boost of velocity $v$.
As sketched in the top panel, the refractive index in the left region is larger than in the right region. This difference in the refractive index is obtained by properly changing the parameters $\beta_i$ and $\Om_i$ appearing in the Lagrangian~\eqref{eq:lagrangian}.
In this plot the values of the velocity $v$ and of the refractive index change $\delta n$ have been arbitrarily chosen for illustrative purposes.
The bottom panels are enlargements of the gray-dot-bordered squared of the respective upper panels. The dispersion relation is graphically solved for a fixed comoving frequency $\om$, chosen in the frequency window in which the black hole horizon is present.
Solutions appear both on the positive-norm positive-$\Om$ branches (solid curves) and on the negative-norm negative-$\Om$ branches (dashed curves).
The open dots denote solutions on the optical branches with positive (o) and negative ($\rm\tilde o$) frequency $\Om$. The arrows indicate the direction of propagation (group velocity in the comoving frame) of the associated modes $V_{\om,L/R}^{\alpha/\tilde\alpha}$. In the left region (left panels), the dispersion relation has only six real-$k$ solutions. 
In the right region (right panels), the real-$k$ solutions are eight, and the two extra solutions are one leftgoing ($V_{\om,R}^{\rm o2}$) and one rightgoing ($V_{\om,R}$).}
\label{fig:dispersion}
\end{figure*}
In Fig.~\ref{fig:dispersion} the full Sellmeier dispersion relation is plotted for $x<0$ (upper left panel) and $x>0$ (upper right panel) in the laboratory reference frame $(\Om,K)$ [see Eq.~\eqref{eq:sellmeier}]. A boost is then performed on the axes and the new axes $(\om,k)$ in the reference frame comoving with the pulse are drawn [see Eq.~\eqref{eq:dispersion}]. The central region of those plots (gray-dot-bordered square) is enlarged in the bottom panels.
The dispersion relation is graphically solved for a fixed value of the comoving frequency $\om$ (dashed line) in the $(\ommin,\ommax)$ range.
There are eight branches: four with positive $\Om$ (solid curves) and four with negative $\Om$ (dashed curves), symmetrically placed in the lower half plane.
In this comoving-frequency range no solution belongs to the highest (positive or negative) energy branches.
We therefore name only the six branches of the dispersion relation with low energy $|\Om|$. In the upper half plane ($\Om>0$), starting from the lowest energy branch we call them lower (l), optical (o), and upper (u). Symmetrically, the three branches with negative laboratory frequency $\Om$ and negative norm (as demonstrated in Sec.~\ref{subsec:matching}) are labeled by $\rm \tilde l$, $\rm \tilde o$, and $\rm \tilde u$.
Accordingly, the solutions of the dispersion relation are labeled by a superscript l, o, u, $\rm \tilde l$, $\rm \tilde o$, and $\rm \tilde u$.
To assist the reader in the comparison with Fig.~\ref{fig:dispsimp}, the solutions on the positive- and negative-frequency optical branches o and $\rm\tilde o$ are denoted by an open dot. The arrow above each solution indicates the direction of propagation (group velocity in the comoving frame) of the associated mode $V_{\om,L/R}^{\alpha/\tilde\alpha}$.

Note that, for $x<0$ (left panels), there are six real-$k$ solutions, all corresponding to leftgoing modes. The remaining two solutions of Eq.~\eqref{eq:dispersion} have complex conjugate $k$. They are associated with exponentially growing ($V_{\om,L}^{\rm grow}$) and decaying ($V_{\om,L}^{\rm dec}$) modes for $x\to-\infty$.
For $x>0$ (right bottom panel), instead, the eight solutions are all real.
The two extra real solutions, that do not have a corresponding solution on the left side, belong to the optical branch and are associated, respectively, with a leftgoing mode, named $V_{\om,R}^{\rm o2}$, and with the unique rightgoing mode, simply named $V_{\om,R}$, without any superscript.

\subsection{Globally defined modes}

In the previous section, plane waves modes propagating in the asymptotically flat left and right regions have been identified.
Combining those asymptotic waves, two relevant bases of globally defined asymptotically bounded modes (GDMs) (not diverging at infinity) can be constructed.

We define the \emph{in} basis as the set of \emph{in} modes, whose asymptotic decomposition~\eqref{eq:leftright} has only one AM with group velocity $v_g$ directed toward $x=0$. We say that the group velocity of an AM is directed toward the horizon if $v_g>0$ ($v_g<0$) for modes which are solutions of the mode equation in the left (right) region.

Analogously, we define the \emph{out} basis as the set of \emph{out} modes, whose asymptotic decomposition has only one AM with group velocity directed toward $x=-\infty$ ($x=+\infty$) if the AM is a solution of the field equation in the left (right) region.

\begin{figure}
 \includegraphics{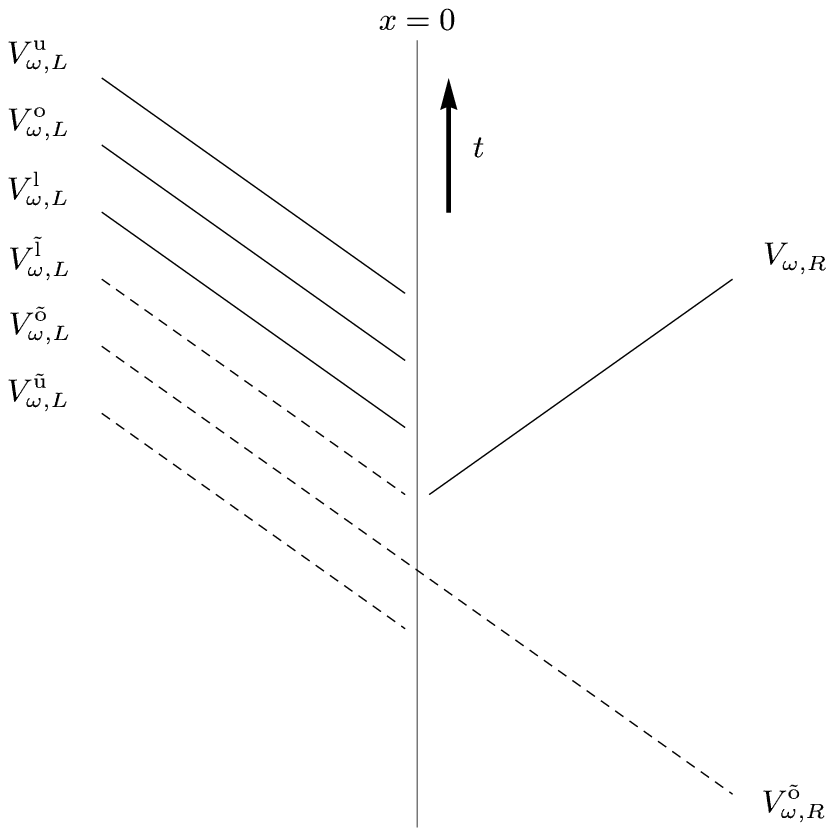}
 \caption{Asymptotic decomposition of the globally defined \emph{in} mode $V_{\omega}^{\rm in,\tilde o}$. The only asymptotic branch with group velocity directed toward the horizon is $V_{\omega,R}^{\tilde o}$. Positive- (negative-) norm modes are represented by solid (dashed) lines.}
 \label{fig:mode}
\end{figure}

To clarify the meaning of these definitions with an example, the asymptotic decomposition of one of the three negative-norm \emph{in} GDMs is schematically represented in Fig.~\ref{fig:mode}. We name it $V_{\omega}^{\rm in,\tilde o}$ since its unique incoming AM is $V_{\omega,R}^{\tilde o}$; that is, in the asymptotic decomposition of Eq.~\eqref{eq:leftright} $R_{\om}^{\tilde o}$ is the only nonvanishing coefficient associated with a mode with group velocity directed toward the horizon (see Fig.~\ref{fig:dispersion}).

\begin{table*}
\begin{ruledtabular}
\begin{tabular}{ccccccccccccccccc}
&\multicolumn{8}{c}{Coefficients of left modes}&\multicolumn{8}{c}{Coefficients of right modes}\\
\cline{2-9}\cline{10-17}\\[-1.1em]
\vspace{0.2em}
& $V_{\omega,L}^{\rm\tilde l}$ & $V_{\omega,L}^{\rm\tilde o}$ & $V_{\omega,L}^{\rm\tilde u}$ & $V_{\omega,L}^{\rm l}$ & $V_{\omega,L}^{\rm o}$ & $V_{\omega,L}^{\rm u}$ & $V_{\omega,L}^{\rm dec}$ & $V_{\omega,L}^{\rm grow}$ 
& $V_{\omega,R}^{\rm\tilde l}$ & $V_{\omega,R}^{\rm\tilde o}$ & $V_{\omega,R}^{\rm\tilde u}$ & $V_{\omega,R}^{\rm l}$ & $V_{\omega,R}^{\rm o}$ & $V_{\omega,R}^{\rm o2}$ & $V_{\omega,R}^{\rm u}$ & $V_{\omega,R}$\\
\hline
$V_{\omega}^{\rm in,\tilde l}$
& $\alpha_\om^{\rm l}$ & $A_\om^{\rm l,o}$ & $A_\om^{\rm l,u}$ & $B_{\omega}^{\rm l}$ & $B_{\omega}^{\rm l,o}$ & $B_{\omega}^{\rm l,u}$ & $D_{\omega}^{\rm l}$ & 0 
& 1 & 0 & 0 & 0 & 0 & 0 & 0 & $\beta^{\rm l}_{\omega}$
\\
$V_{\omega}^{\rm in,\tilde o}$
& $A_{\omega}^{\rm o,l}$ & $\alpha_\om$ & $A_\om^{\rm o,u}$ & $B_{\omega}^{\rm o,l}$ & $B_{\omega}$ & $B_{\omega}^{\rm o,u}$ & $D_{\omega}^{\rm o}$ & 0 
& 0 & 1 & 0 & 0 & 0 & 0 & 0 & $\beta_\om$
\\
$V_{\omega}^{\rm in,\tilde u}$
& $A_\om^{\rm u,l}$ & $A_\om^{\rm u,o}$ & $\alpha_\om^{\rm u}$ &  $B_{\omega}^{\rm u,l}$ & $B_{\omega}^{\rm u,o}$ & $B_{\omega}^{\rm u}$ & $D_{\omega}^{\rm u}$ & 0 
& 0 & 0 & 1 & 0 & 0 & 0 & 0 & $\beta_\omega^{\rm u}$
\end{tabular}
\end{ruledtabular}
\caption{Coefficients of the asymptotic decomposition of the globally defined \emph{in} modes $V_{\omega}^{\rm in,\tilde l}$, $V_{\omega}^{\rm in,\tilde o}$, and $V_{\omega}^{\rm in,\tilde u}$ on the bases of left and right AMs.}
\label{tab:mode}
\end{table*}
In Table~\ref{tab:mode}, all the coefficients $L_\om^\alpha$'s and $R_\om^\alpha$'s are reported for the asymptotic decompositions of the three negative-norm \emph{in} GDMs $V_{\omega}^{\rm in,\tilde l}$, $V_{\omega}^{\rm in,\tilde o}$, $V_{\omega}^{\rm in,\tilde u}$.
As noticed in the previous section, both the AMs $V_{\om,R}^{\rm o2}$ and $V_{\om,R}$, which are present only in the right region, lie on the optical branch of the dispersion relation.
As mentioned at the beginning of Sec.~\ref{subsec:asymptotic}, in the simplified analysis restricted to the optical branch, analog Hawking radiation is expected on the AM $V_{\om,R}$, due to the scattering at the analog horizon of the optical negative-norm \emph{in} GDM $V_{\omega}^{\rm in,\tilde o}$.
Thus, aiming to extend the definition of Hawking radiation to the present situation, adopting standard notation, we name $\beta_\om$ the coefficient of $V_{\om,R}$ in the expansion of $V_{\omega}^{\rm in,\tilde o}$.
Similarly, we name $\alpha_\om$ the coefficient of $V_{\omega,L}^{\rm\tilde o}$.
Furthermore, adopting the notation of Ref.~\cite{MacherRP1}, the coefficients of modes with negative (positive) norm, in the expansion of modes with negative norm are named $A$ ($B$).
As an example, in the expansion of the negative-norm mode $V_{\omega}^{\rm in,\tilde o}$, the coefficients of the negative-norm modes $V_{\omega,L}^{\rm \tilde l}$ and $V_{\omega,L}^{\rm \tilde u}$ are named $A_\om^{\rm o,l}$ and $A_\om^{\rm o,u}$. The coefficients of the positive-norm modes $V_{\omega,L}^{\rm\tilde o}$, $V_{\omega,L}^{\rm\tilde l}$, and $V_{\omega,L}^{\rm\tilde u}$ are named $B_\om$, $B_\om^{\rm o,l}$, and $B_\om^{\rm o,u}$, respectively.

Finally, the eight unknown coefficients in the two left and right asymptotic expansions for each GDM are determined by imposing the matching conditions (eight independent equations), derived in Sec.~\ref{subsec:matching}.

By proceeding along this line, it is possible to construct seven \emph{in} GDMs---$V_{\omega}^{\rm in,\tilde l}$, $V_{\omega}^{\rm in,\tilde o}$, $V_{\omega}^{\rm in,\tilde u}$, $V_{\omega}^{\rm in,l}$, $V_{\omega}^{\rm in,o}$, $V_{\omega}^{\rm in,u}$, $V_{\omega}^{\rm in,o2}$---whose unique branches with group velocity directed toward the horizon are, respectively, $V_{\omega,R}^{\rm \tilde l}$, $V_{\omega,R}^{\rm \tilde o}$, $V_{\omega,R}^{\rm \tilde u}$, $V_{\omega,R}^{\rm l}$, $V_{\omega,R}^{\rm o}$, $V_{\omega,R}^{\rm u}$, $V_{\omega,R}^{\rm o2}$.
Similarly, one constructs seven \emph{out} GDMs---$V_{\omega}^{\rm out,\tilde l}$, $V_{\omega}^{\rm out,\tilde o}$, $V_{\omega}^{\rm out,\tilde u}$, $V_{\omega}^{\rm out,l}$, $V_{\omega}^{\rm out,o}$, $V_{\omega}^{\rm out,u}$, $V_{\omega}^{\rm out}$---whose unique branches with group velocity directed to infinity are, respectively, $V_{\omega,L}^{\rm \tilde l}$, $V_{\omega,L}^{\rm \tilde o}$, $V_{\omega,L}^{\rm \tilde u}$, $V_{\omega,L}^{\rm l}$, $V_{\omega,L}^{\rm o}$, $V_{\omega,L}^{\rm u}$, $V_{\omega,R}$.

\subsection{The scattering matrix}

As in Eq.~\eqref{eq:expansionadagger}, the field operator $V$ is expanded indifferently with respect either to the \emph{in} or to the \emph{out} basis of GDMs,
\begin{align}\label{eq:expansiondagger}
 V
 &=\int_0^\infty\!\!\!\dd
\om\,\ee^{-\ii\om t}\left(\sum_{\alpha\in P}  V_{\om}^{{\rm in},\alpha}{\hat a}_{\om}^{{\rm in},\alpha}
+\sum_{\tilde\alpha\in N} V_{\om}^{{\rm in},\tilde\alpha}
{\hat a}_{\om}^{{\rm in},\tilde\alpha\dagger}\right)\nonumber\\
&\qquad\qquad
+\mbox{H.c.}\\
&=\int_0^\infty\!\!\!\dd
\om\,\ee^{-\ii\om t}\!\left(\sum_{\alpha\in P} V_{\om}^{{\rm out},\alpha}{\hat a}_{\om}^{{\rm out},\alpha}
+\!\sum_{\tilde\alpha\in N} V_{\om}^{{\rm out},\tilde\alpha}
{\hat a}_{\om}^{{\rm out},\tilde\alpha\dagger}\right)
\nonumber\\&\qquad\qquad
+\mbox{H.c.}.
\end{align}
The transformation between the two bases follows straightforwardly from the construction of the previous section. 
The coefficients of the matrix connecting the \emph{in} and \emph{out} orthogonal bases can be directly read from the expansion of \emph{in} and \emph{out} modes on the orthogonal bases of left and right AMs.
For instance, from the second line of Table~\ref{tab:mode},
\begin{multline}
 V_{\omega}^{\rm in,\tilde o}=
A_{\omega}^{\rm o,l}V_{\omega}^{\rm out,\tilde l}+
\alpha_\om V_{\omega}^{\rm out,\tilde o}+
A_\om^{\rm o,u}V_{\omega}^{\rm out,\tilde u}
+B_{\omega}^{\rm o,l} V_{\omega}^{\rm out,l}\\
+B_{\omega} V_{\omega}^{\rm out,o}+
B_{\omega}^{\rm o,u} V_{\omega}^{\rm out,u}+
\beta_\om V_{\omega}^{\rm out}.
\end{multline}
Repeating this procedure for each \emph{in} mode, the scattering matrix $S$ is fully determined
\begin{equation}
 V_{\om}^{\rm in,\beta}=\sum_{\beta'} S^{\beta\beta'}V_{\om}^{\rm out,\beta'},
\end{equation}
where $\beta$ and $\beta'$ run over all positive- and negative- norm modes. The relation between \emph{in} and \emph{out} destruction and creation operators is easily derived from the $S$ matrix,
\begin{equation}\label{eq:ST}
 \hat A^{\rm out}=S^T \hat A^{\rm in},
\end{equation}
where
\begin{align}
 &\hat A^{\rm in} =
 \begin{pmatrix}
  {\hat a}_{\omega}^{{\rm in,\tilde l}\,\dagger} & {\hat a}_{\omega}^{{\rm in,\tilde o},\,\dagger} & {\hat a}_{\omega}^{{\rm in,\tilde u}\,\dagger} & {\hat a}_{\omega}^{\rm in,l} & {\hat a}_{\omega}^{\rm in,o} & {\hat a}_{\omega}^{\rm in,u} & {\hat a}_{\omega}^{\rm in,o2}
 \end{pmatrix}^T,\\
 &\hat A^{\rm out} \!=\!
 \begin{pmatrix}
  {\hat a}_{\omega}^{{\rm out,\tilde l}\,\dagger}\! & {\hat a}_{\omega}^{{\rm out,\tilde o}\,\dagger}\! & {\hat a}_{\omega}^{{\rm out,\tilde u}\,\dagger}\! & {\hat a}_{\omega}^{\rm out,l}\! & {\hat a}_{\omega}^{\rm out,o}\! & {\hat a}_{\omega}^{\rm out,u}\! & {\hat a}_{\omega}^{\rm out}
 \end{pmatrix}\!^T
\end{align}
are seven-dimensional vector formed, respectively, by the \emph{in} and \emph{out}
creation operators ${\hat a}_{\om}^{{\rm in},\tilde\alpha\dagger}$ and ${\hat a}_{\om}^{{\rm out},\tilde\alpha\dagger}$, respectively, of the negative-norm modes,
and by the \emph{in} and \emph{out} destruction operators ${\hat a}_{\om}^{{\rm in},\alpha}$ and ${\hat a}_{\om}^{{\rm out},\alpha}$, respectively, of the positive-norm modes.

\begin{figure*}
 \includegraphics{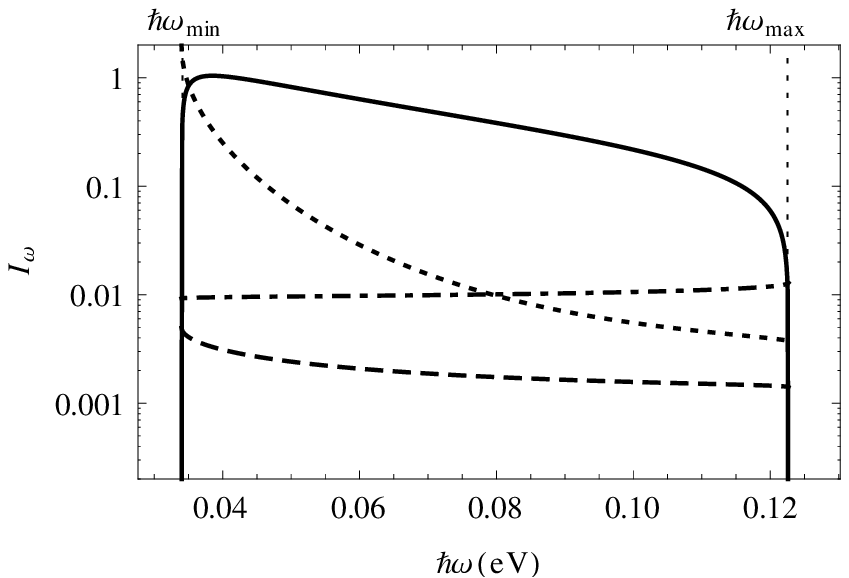}
 \hspace{1.3em}
 \includegraphics{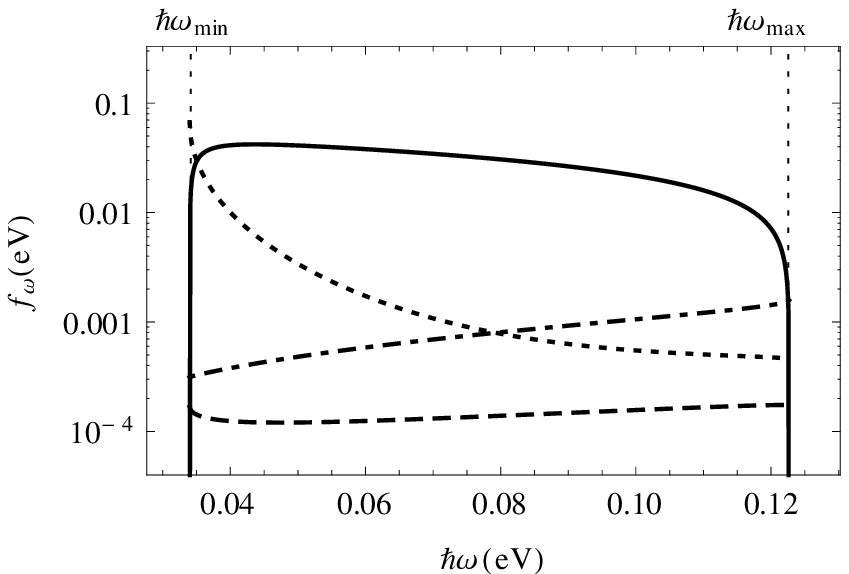}
 \caption{{\it Black-hole-like configuration.} Particle fluxes (left panel) $I_\om^{\rm out}$ (solid line), $I_\om^{\rm out,l}$ (dashed line), $I_\om^{\rm out,o}$ (dotted line), and $I_\om^{\rm out,u}$ (dot-dashed line) and the corresponding energy fluxes (right panel) $f_\om^{\rm out}$, $f_\om^{\rm out,l}$, $f_\om^{\rm out,o}$, and $f_\om^{\rm out,u}$, as seen from the comoving reference frame, for a black hole horizon configuration, as sketched in Fig.~\ref{fig:dispersion}. Parameters $v=0.66c$ and $\epsilon=0.3$, yielding via Eq.~\eqref{eq:betaomleft} a large refractive index jump of $\delta n=0.12$ for optical frequencies.}
 \label{fig:flux}
\end{figure*}
%

\section{Spontaneous emission}
\label{sec:emission}

In this section we make use of the $S$ matrix calculated in the previous section to provide quantitative predictions for some of the simplest observable quantities such as the intensity and the spectrum of the quantum vacuum emission. A straightforward extension of the formalism following the lines of~\cite{Recati2009,MacherBEC} can be used to obtain information on more sophisticated quantities like the correlation properties of the quantum vacuum emission into the different outgoing modes: Given the importance of correlation experiments to assess the quantum vacuum nature of the emission, this problem will be the subject of future work.

\subsection{Comoving frame}

For the sake of simplicity, we make the reasonable assumption that there are no ingoing particles; that is, the system is in the vacuum state defined by the destruction operators
associated with the \emph{in} modes:
\begin{equation}
 \hat a_\om^{{\rm in},\alpha}|0_{\rm in}\rangle=0,\quad \hat a_{\om}^{{\rm in},\tilde\alpha}|0_{\rm in}\rangle=0.
\end{equation}
The occupation numbers of \emph{out} modes on this state are easily computed using Eq.~\eqref{eq:ST}. They, in general, do not vanish, since Eq.~\eqref{eq:ST} mixes creation and destruction operators. For instance, the expected occupation number of the unique rightgoing mode $V_\om^{\rm out}$ on the state $|0_{\rm in}\rangle$ is
\begin{equation}
 \langle0_{\rm in}| 2\pi\hat a_\om^{\rm out\dagger}\hat a_\om^{\rm out}|0_{\rm in}\rangle
 =2\pi\delta(0)(|\beta_\om|^2+|\beta_\om^{\rm l}|^2+|\beta_\om^{\rm u}|^2),
\end{equation}
where the factor $2\pi$ has been inserted coherently with the normalization of the Fock operators of Eq.~\eqref{eq:fockcomm}.
As usual, there is an infrared divergence associated with the quantization of a field theory in an infinite space-time volume. Going to a finite-size time box $\Delta t$, one must replace
\begin{equation}
 2\pi\delta(\om-\om')\longrightarrow \Delta t\,\delta_{\om\om'}.
\end{equation}
This implies that the number of particles $\Delta n$ created in a time $\Delta t$ at a frequency $\om$ is
\begin{equation}
 \Delta n_\om^{\rm out}
 =(|\beta_\om|^2+|\beta_\om^{\rm l}|^2+|\beta_\om^{\rm u}|^2)\Delta t.
\end{equation}

Analogously, there are several other channels in which particles are created, related to the mixing of the other positive- and negative-norm modes. For the positive-frequency modes one obtains
\begin{align}
 \Delta n_\om^{\rm out,l}&=(|B_\om^{\rm l}|^2+|B_\om^{\rm o,l}|^2+|B_\om^{\rm u,l}|^2)\Delta t,\\
 \Delta n_\om^{\rm out,o}&=(|B_\om|^2+|B_\om^{\rm l,o}|^2+|B_\om^{\rm u,o}|^2)\Delta t,\\
 \Delta n_\om^{\rm out,u}&=(|B_\om^{\rm u}|^2+|B_\om^{\rm l,u}|^2+|B_\om^{\rm o,u}|^2)\Delta t.
\end{align}

Since in the comoving reference frame the source of photons (the pulse) is at rest, the number of created particles $r_\om$ per unit time and unit bandwidth coincides with the flux of particles $I_\om$ crossing a certain surface at constant $x$.
Thus, the flux of particles per unit time and unit bandwidth in the comoving reference frame is
\begin{equation}\label{eq:Iom}
 I_\om^{\rm out,\alpha}=r_\om^{\rm out,\alpha}=\frac{\dd n^{\rm out,\alpha}}{\dd t\,\dd\om}=\frac{\Delta n_\om^{\rm out,\alpha}}{\Delta t}.
\end{equation}
The flux of energy (which coincides with the energy production rate in the reference frame where the source is at rest) associated with the mode $\alpha$ is
\begin{equation}
 f_\om^{\rm out,\alpha}=\frac{\dd E^{\rm out,\alpha}}{\dd t\,\dd\om}=\hbar\om\, I_\om^{\rm out,\alpha}.
\end{equation}
In Fig.~\ref{fig:flux} we plot the fluxes of particles $I_\om^{\rm out}$ (solid line), $I_\om^{\rm out,l}$ (dashed line), $I_\om^{\rm out,o}$ (dotted line), $I_\om^{\rm out,u}$ (dot-dashed line) and the respective energy fluxes (right panel) $f_\om^{\rm out}$, $f_\om^{\rm out,l}$, $f_\om^{\rm out,o}$, and $f_\om^{\rm out,u}$, for a pulse moving with velocity $v=0.66c$. The values of $\beta_{i,R}$ and $\Om_{i,R}$ in the right region have been chosen accordingly to the dispersion relation in fused silica, the material used in the experiment of Ref.~\cite{faccioexp}:
\begin{equation}
\begin{array}{ll}
 \beta_{1,R}=0.071\,419\,14,\quad &\hbar\Om_{1,R}=0.125\,285\,\mbox{eV},\\
 \beta_{2,R}=0.032\,463\,04,\quad &\hbar\Om_{2,R}=10.6661\,\mbox{eV},\\
 \beta_{3,R}=0.055\,399\,15,\quad &\hbar\Om_{3,R}=18.1252\,\mbox{eV}.
\end{array}
\end{equation}
In the left region, for illustrative purposes we used
\begin{equation}\label{eq:betaomleft}
 \beta_{i,L}=(1+\epsilon)\beta_{i,R},\quad \Om_{i,L}=(1+\epsilon)^{-1/2}\Om_{i,R},
\end{equation}
with a very large value of $\epsilon=0.3$. When $\Om$ is in the optical range and far enough from the poles of the dispersion relation
\begin{equation}\label{eq:deltan}
 \delta n= n_L-n_R\approx\frac{n_R^2-1}{2n_R}\epsilon,
\end{equation}
that yields a quite large value of $\delta n\approx 0.12$, which requires a very strong laser intensity $I\approx3\times10^{14}\,\mbox{W}/\mbox{cm}^2$~\cite{faccioexp,kinetics}.

\begin{figure}
  \includegraphics{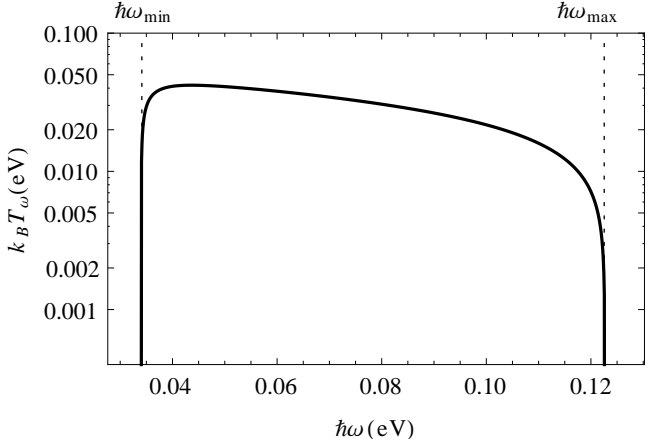}
 \caption{{\it Black-hole-like configuration.} Frequency-dependent temperature $T_\om$ of the Hawking-like radiation $I_\om^{\rm out}$, as defined in Eq.~\eqref{eq:tom}. 
The emission spectrum is observed in the comoving reference frame for the black hole horizon configuration considered in Fig.~\ref{fig:flux}.}
 \label{fig:tom}
\end{figure}

From Fig.~\ref{fig:flux}, it is immediate to see that the dominant contribution to the energy flux comes from the rightgoing modes via Hawking-like processes as discussed in the previous section and in Ref.~\cite{kinetics}. However, what can we say about its thermal properties?
It is well known that, in the $1+1$ dimensional case, the energy flux of a thermal source goes to a constant value in the low-frequency limit $\om\to0$. In the present situation, however, the Hawking-like channel is open only in a narrow range of frequencies $\ommin<\om<\ommax$ (see Fig.~\ref{fig:dispsimp}), and the low-frequency limit cannot be taken: As a result, asking whether the spectrum is thermal or not is not really a well-posed question. Nevertheless, one can observe that the energy flux does not vary much within the $\ommin<\om<\ommax$ frequency window; and exception is made for frequencies in the close vicinity of the boundaries. To make this statement more quantitative, in Fig.~\ref{fig:tom} we plot the (frequency-dependent) temperature $T_\om$ corresponding to the rightgoing flux $I_\om^{\rm out}$, as  implicitly defined by
\begin{equation}\label{eq:tom}
 I_\om^{\rm out}=\frac{1}{\ee^{\hbar\om/k_B T_\om}+1}.
\end{equation}
If the spectrum were perfectly thermal, $T_\om$ would be independent of frequency. In the present situation, even if the value of $T_\om$ maintains the same order of magnitude for frequencies within the $(\ommin,\ommax)$ frequency range, the spectrum is still substantially different from a thermal one.

\begin{figure}[b]
 \includegraphics{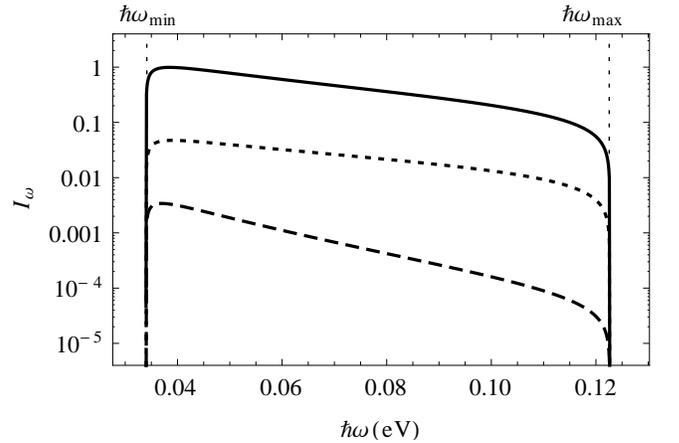}
 \caption{{\it Black-hole-like configuration.} Separate contribution of $|\beta_\om|^2$ (solid line), $|\beta_\om^{\rm l}|^2$ (dashed line), and $|\beta_\om^{\rm u}|^2$ (dotted line) to the flux of particles $I_\om^{\rm out}$ for the black hole horizon configuration considered in Fig.~\ref{fig:flux}.}
 \label{fig:rightgoing}
\end{figure}

To investigate more deeply the features of the spectrum of $V_\om^{\rm out}$ (the global modes associated with the asymptotic plane wave $V_{\om,R}$; see Fig.~\ref{fig:dispersion}), in Fig.~\ref{fig:rightgoing} the flux of particles is separated into its components due to $|\beta_\om|^2$ (solid line), $|\beta_\om^{\rm l}|^2$ (dashed line), and $|\beta_\om^{\rm u}|^2$ (dotted line).
Note that the contributions to spontaneous emission on the mode $V_\om^{\rm out}$ by the lower and the upper branch are negligible with respect to the term coming from the optical negative-frequency mode $V_{\om}^{\rm in,\tilde o}$, associated with the AM $V_{\om,R}^{\rm\tilde o}$ of Fig.~\ref{fig:dispersion}. This behavior was anticipated in Ref.~\cite{kinetics} and it has been now confirmed by the present calculation.

To make our discussion complete, it is important to assess whether the full three-pole Sellmeier dispersion relation is really needed or simplified models based on modified dispersion relations are enough to properly reproduce the main features of the emission.
As a relevant example, we consider a single-pole form of the dispersion relation, for which the optical branch (in this case the lower polariton) has a simple subluminal dispersion.
The details of the calculation are given in Appendix~\ref{app:thermal}: The main conclusion is that, in contrast to the full Sellmaier dispersion, in this case the emission spectrum turns out to be perfectly thermal from arbitrarily low frequencies up to a certain cutoff frequency.
To understand this result, it is enough to note that this simplified dispersion admits an analog geometry in the low-frequency limit, so that the concept of horizon is well defined in the standard gravitational sense.
Any material where the low-frequency pole is not present, such as diamond~\cite{refractiveindex}, would therefore show a thermal spectrum.
On the other hand, in the case of the full Sellmeier dispersion, the dramatic deviation of the emission spectrum from thermality stems from the presence of the low-frequency pole in the infrared region.

\subsection{Laboratory frame}

In the previous section we investigated the spectral properties of the quantum vacuum emission as observed from the frame comoving with the pulse; in particular, we have provided a critical comparison with the thermal spectrum that is expected for standard Hawking radiation. In this section, we show how an even more dramatic departure from a thermal spectrum is obtained when the emission is observed in the laboratory reference frame.

With an eye to recent experimental studies, we focus our attention on the total number of particles created by the moving pulse rather than the flux of particles per unit time. In fact, the pulse stably propagates only for a very short distance $\Delta X_s$ (about 1 mm~\cite{faccionjp}), corresponding to a duration of $\Delta T_s=\Delta X_s/V_g$. In an experiment one observes the pulse in this small region and measures the photons produced therein. Naming $R_\Om$ the production rate of particles at frequency $\Om$, as measured in the laboratory, the number of particles produced by the pulse in the laboratory time $\dd T$ in the range of frequency ($\Om$,$\Om+\dd\Om$) is
\begin{equation}
 \Delta N=R_\Om\dd T\dd\Om.
\end{equation}
Since the number of created particles is invariant under Lorentz transformation,
\begin{equation}
 \Delta N=r_\Om\dd t\dd\om;
\end{equation}
that is, $\Delta N$ can also be computed as the number of particles created in the comoving time $\dd t$ by the pulse, in the frequency range $(\om,\om+\dd\om)$. To compute the relation between $R_\Om$ and $r_\om$, it is enough to compute the transformation rules of the frequency range and of the time interval.
Using the inverse of the Lorentz transformation~\eqref{eq:boosfrequency}
\begin{equation}
 \om=\gamma(\Om-v K),
\end{equation}
$\dd\om$ can be expressed in term of $\dd\Om$ as
\begin{equation}
 \dd\om=\gamma\left(\dd\Om-v\frac{\dd K}{\dd\Om}\dd\Om\right)=\gamma\left(1-\frac{v}{V_g(\Om)}\right)\dd\Om,
\end{equation}
where $V_g$ is the particle group velocity measured in the laboratory reference frame.
The transformation of the time interval is found by noting that the source (the pulse) is at rest in the comoving frame, so $\dd t$ is its proper time interval. Consequently, the corresponding laboratory time interval $\dd T$ is given by the usual Lorentz dilation of time,
\begin{equation}
 \dd T=\gamma\dd t.
\end{equation}
Putting everything together, the rate of particle production, as seen in the laboratory frame is
\begin{equation}
 R_\Om=\left(1-\frac{v}{V_g}\right)r_\om.
\end{equation}

The corresponding energy production rate $\varepsilon_\Om$ in the laboratory frame is
\begin{equation}
 \varepsilon_\Om=\hbar\Om R_\Om = \hbar\Om I_\Om\left(1-\frac{v}{V_g}\right).
\end{equation}
\begin{figure}
 \includegraphics{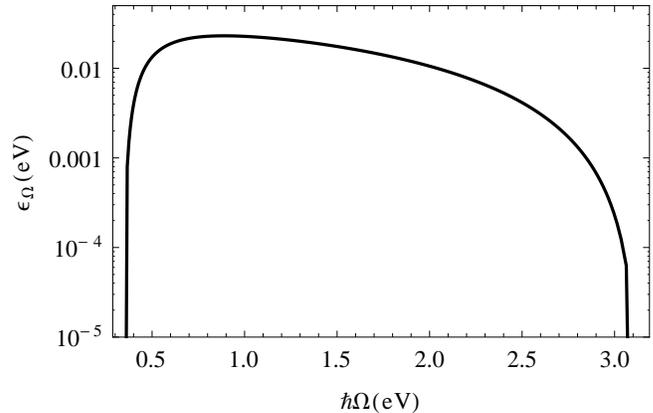}
 \caption{{\it Black-hole-like configuration.} Energy production rate $\varepsilon_\Om^{\rm out}$ on the rightgoing mode $V_{\om(\Om)}^{\rm out}$ as seen by an observer in the laboratory, for the black hole horizon configuration considered in Fig.~\ref{fig:flux}.}
 \label{fig:fluxlab}
 \end{figure}
This quantity is plotted in Fig.~\ref{fig:fluxlab} for the rightgoing mode $V_\om^{\rm out}$, which is responsible for the Hawking-like emission.
The effect of the boost is twofold. First, it shifts the emission frequency in the laboratory: $\Om$ is much larger than the comoving frequency $\om$, reaching optical frequencies.
Unfortunately, as observed in Ref.~\cite{kinetics}, this boost effect is completely lost in a three-dimensional system when one looks perpendicularly to the pulse propagation. Radiation at such high frequency could be observed perpendicularly only if some scattering process changed the direction of photons.

\begin{figure*}
 \includegraphics{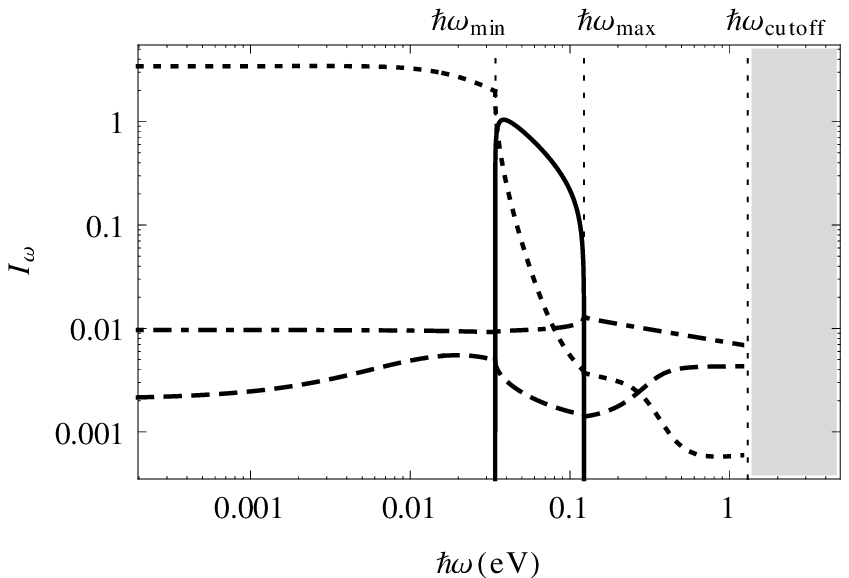}
 \hspace{1.3em}
 \includegraphics{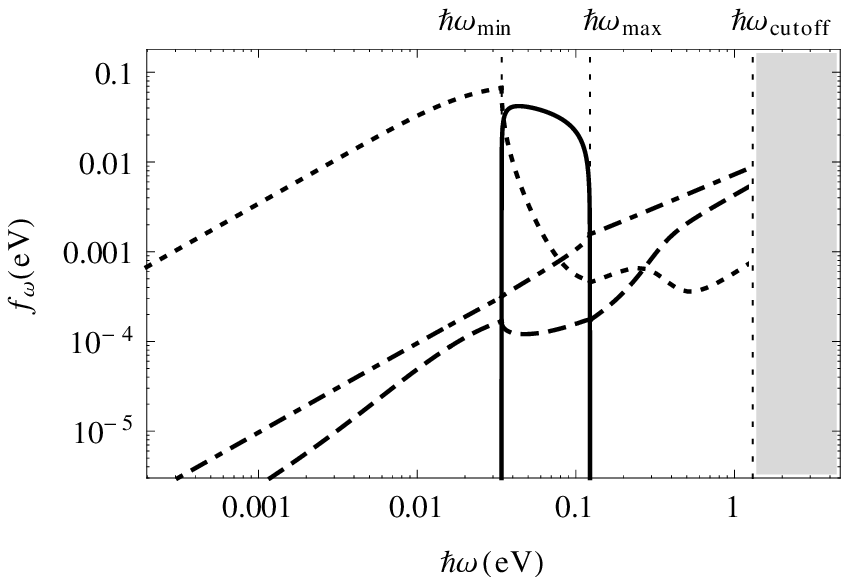}
 \caption{{\it Black-hole-like configuration.} Fluxes of particles $I_\om^{\rm out}$ (solid line), $I_\om^{\rm out,l}$ (dashed line), $I_\om^{\rm out,o}$ (dotted line), and $I_\om^{\rm out,u}$ (dot-dashed line) and the respective energy fluxes (right panel) $f_\om^{\rm out}$, $f_\om^{\rm out,l}$, $f_\om^{\rm out,o}$, and $f_\om^{\rm out,u}$ for the black hole horizon configuration considered in Fig.~\ref{fig:flux}.
 The plot has been restricted to $\om<\om_{\rm cutoff}$. $I_\om^{\rm out}$ and $f_\om^{\rm out}$ (solid lines) are defined only for $\ommin<\om<\ommax$.}
 \label{fig:fluxall}
\end{figure*}

To conclude this section, it is interesting to give a rough estimate of the number of photons emitted on the rightgoing mode $V^{\rm out}_{\om(\Om)}$ by this process. 
Integrating over frequency the production rate in the laboratory frame and multiplying it by the time $\Delta T_s=\Delta X_s/V$ over which the pulse stably propagates, we obtain (for $\Delta X_s\approx 1\,\mbox{mm}$)
\begin{equation}
 N_{\rm pulse}=\int{R_\Om}\dd\Om\,\frac{\Delta X_s}{V}\approx240.
\end{equation}
Of course, the same result can be obtained by directly integrating the particle production rate $r_\om$, measured in the comoving frame (coinciding with the flux $I_\om$), over the comoving frequency $\om$ and multiplying by the proper duration of the pulse $\Delta t_s=\Delta T_s/\gamma(V)$.

\subsection{Outside the analog horizon frequency window}
\label{subsec:abovebelow}

In the previous subsection we focused on the frequency range where Hawking-like emission is present.
We now investigate the range of comoving frequencies below $\ommin$ and above $\ommax$, where the main particle production channel disappears.
Particle creation is indeed possible also outside the $(\ommin,\ommax)$ range through the other channels, and the dashed, dotted, and dot-dashed lines of Fig.~\ref{fig:flux} can then be extended for $\om<\ommin$ and $\om>\ommax$.
Since for those values of the comoving frequency there are only six real-$k$ solutions also for $x>0$, a new mode analysis must be performed following Sec.~\ref{sec:modes} to compute a new $6\times6$ scattering matrix.
In Fig.~\ref{fig:fluxall} the numerical results of this computation are reported, together with the results for $\ommin<\om<\ommax$, obtained in the previous section.
The flux of particles in the comoving frame $I_\om^{\rm out,l}$ (dashed line), $I_\om^{\rm out,o}$ (dotted line), and $I_\om^{\rm out,u}$ (dot-dashed line) are plotted in the left panel, and the respective energy fluxes $f_\om^{\rm out}$, $f_\om^{\rm out,l}$, $f_\om^{\rm out,o}$ are plotted in the right one.

First, note that the dominant contribution to the emission comes from the frequency range $\ommin<\om<\ommax$, thanks to the presence of the Hawking channel (solid line). This result confirms the naive expectation that the presence of a horizon should enhance the production of particles.

Second, both the occupation numbers and the energy fluxes are continuous at $\om=\ommin$ and $\om=\ommax$, for modes that do not feel the presence of any horizon, that is, when no turning point is present. For these modes the transition between the two regimes is continuous.

Third, it is worth remarking that the high-frequency region of this plot must be read {\it cum grano salis}. Indeed, we used step functions to describe the spatial behavior of $\beta_i$ and $\Omega_i$, such that arbitrary large frequency/momentum modes are excited.
However, in real physical situations, the transition between the two regions $x<0$ and $x>0$ takes place on a finite length and only modes up to a certain frequency $\om_{\rm cutoff}$ are excited.
Thus, Fig.~\ref{fig:fluxall} provides reliable results only up to $\om_{\rm cutoff}$. Beyond this frequency, the off-diagonal coefficients of the scattering matrix $S$ go to zero exponentially with $\om$ and $S$ reduces to the identity.
Indeed, for $\om>\om_{\rm cutoff}$, modes do not mix because they are well approximated by their WKB expansion.
In a realistic situation, $\om_{\rm cutoff}$ is not larger than the frequency associated with the width of the steepening of the pulse~\cite{faccionjp,angus} (approximately $1~\mu\mbox{m}$), corresponding to $\om_{\rm cutoff}\approx 2\times 10^{15}\,\mbox{s}$ ($\hbar\om_{\rm cutoff}\approx1.3\,\mbox{eV}$).

\section{Particle production in other configurations}
\label{sec:otherconf}

In the previous section, we derived the flux of spontaneously created particles in a configuration closely resembling a black hole geometry.
However, as pointed out in Ref.~\cite{kinetics}, radiation from vacuum fluctuation is expected in optical systems even in the absence of horizons, provided that negative-norm modes with positive comoving frequency are present. We now apply the techniques introduced in this paper to a horizonless configuration for a weaker perturbation ($\delta n\approx 0.001$), similar to the one experimentally realized in Ref.~\cite{faccioexp}.
As a preliminary step, however, we compute the emitted flux for a horizon configuration with a value of $\delta n \approx 0.001$, which coincide with the experimental value~\cite{faccioexp}, so that it is possible to compare the flux produced by a weak perturbation, both in the presence and in the absence of horizons.

\subsection{Small refractive index jump}
\label{subsec:smalln}

\begin{figure}
 \includegraphics{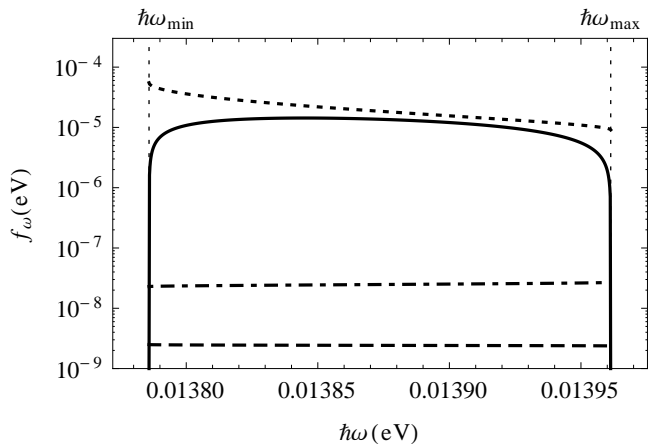}
 \caption{{\it Black-hole-like configuration---small $\delta n$.} Energy fluxes $f_\om^{\rm out}$ (solid line), $f_\om^{\rm out,l}$ (dashed line), $f_\om^{\rm out,o}$ (dotted line), and $f_\om^{\rm out,u}$ (dot-dashed line), for a configuration with a small refractive index jump ($\delta n\approx0.001$), obtained using $\epsilon=0.0026$ in Eq.~\eqref{eq:betaomleft}. The pulse velocity has been fine tuned to $v=0.6838c$ to have an analog black hole horizon.}
 \label{fig:flux_smalln}
\end{figure}

In a realistic situation $\delta n$ is generally a couple of orders of magnitude smaller than the value used in the previous section.
Unfortunately, if $\delta n$ is small,
the pulse velocity $v$ must be extremely fine tuned in order to obtain a configuration with a horizon, as in Fig.~\ref{fig:dispersion}. At the same time, the frequency window $(\ommin,\ommax)$ becomes very narrow and the emission on the rightgoing mode $V_\om^{\rm out}$ is strongly suppressed. In Fig.~\ref{fig:flux_smalln}, the energy fluxes associated with the spontaneous particle production are represented for $\delta n=0.001$. This value has been obtained by modifying the values of $\beta_i$ and $\Om_i$ in the left region, as in Eq.~\eqref{eq:betaomleft}, with $\epsilon\approx 0.0026$. The pulse velocity is $v=0.6838c$, $\hbar\ommin=0.013\,79\,\mbox{eV}$, and $\hbar\ommax=0.013\,96\,\mbox{eV}$.

\begin{figure*}
 \includegraphics{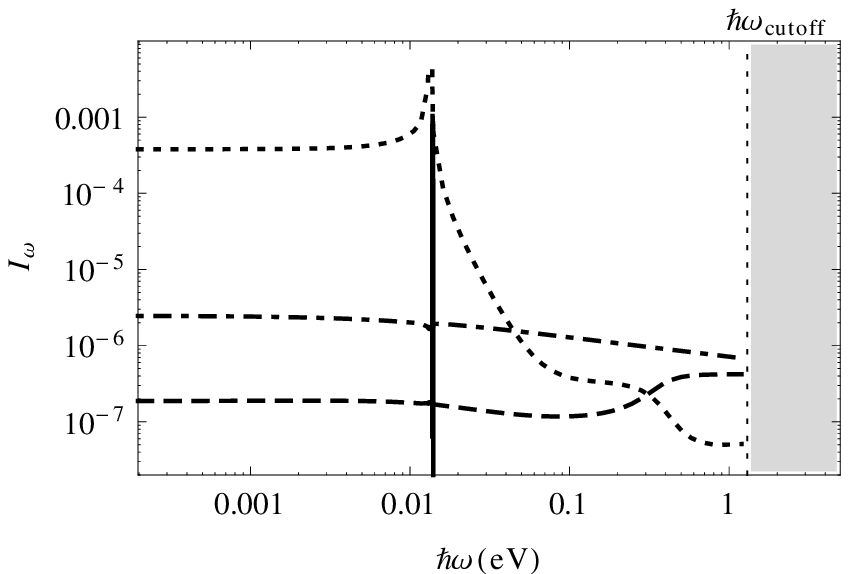}
 \hspace{1.3em}
 \includegraphics{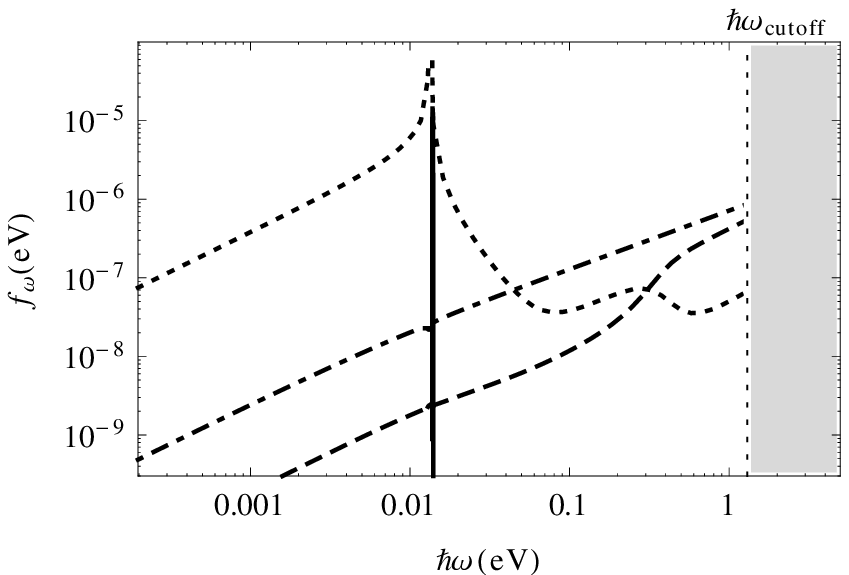}
 \caption{{\it Black-hole-like configuration---small $\delta n$.} Fluxes of particles $I_\om^{\rm out}$ (solid line), $I_\om^{\rm out,l}$ (dashed line), $I_\om^{\rm out,o}$ (dotted line), and $I_\om^{\rm out,u}$ (dot-dashed line) and the respective energy fluxes $f_\om^{\rm out}$, $f_\om^{\rm out,l}$, $f_\om^{\rm out,o}$, and $f_\om^{\rm out,u}$, for the black hole horizon configuration considered in Fig.~\ref{fig:flux_smalln}.
The plot has been restricted to $\om<\om_{\rm cutoff}$. $I_\om^{\rm out}$ and $f_\om^{\rm out}$ (solid lines) are defined only for $\ommin<\om<\ommax$.}
 \label{fig:fluxall_smalln}
\end{figure*}

In Fig.~\ref{fig:fluxall_smalln}, the fluxes of particles and the respective energy fluxes are plotted for any frequency smaller than the cutoff $\om_{\rm cutoff}$, arbitrarily chosen at $\hbar\om_{\rm cutoff}\approx 1.3\,\mbox{eV}$.
The Hawking channel is no longer the dominant one, but its flux is now comparable with the flux of leftgoing particles on the optical-branch mode $V_\om^{\rm out,o}$.
Note also that the range $(\ommin,\ommax)$ is so narrow to be almost invisible on the scale of the figure. As a consequence its contribution to the total flux of leftgoing particles is very small.
\begin{figure}
 \includegraphics{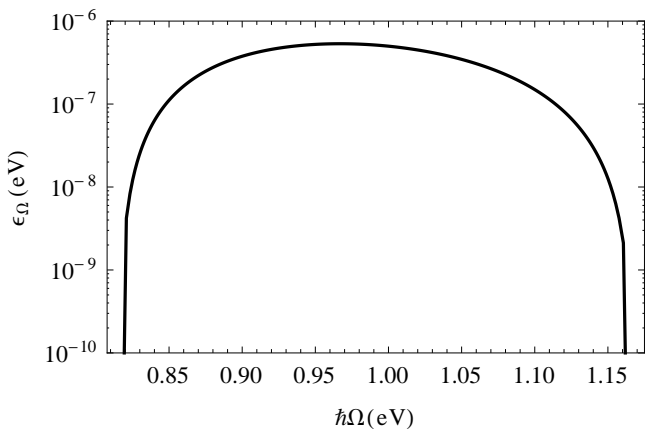}
 \caption{{\it Black-hole-like configuration---small $\delta n$.} Energy production rate $\varepsilon_\Om^{\rm out}$ of the rightgoing mode $V_{\om(\Om)}^{\rm out}$, as seen by an observer in the laboratory, for the black hole horizon configuration considered in Fig.~\ref{fig:flux_smalln}.}
 \label{fig:fluxlab_smalln}
\end{figure}

Furthermore, the number of particles produced on the lower and the upper branches are about three orders of magnitude smaller than the production on the optical branch (both right- and leftgoing). We have also checked that the main contribution to particle production comes from the incoming negative-frequency mode on the optical branch.
Both those results confirm that the lower and upper branch can be safely neglected, as anticipated in Ref.~\cite{kinetics}. Note that the latter result was already visible in Fig.~\ref{fig:rightgoing}, for a different configuration with a larger refractive index jump.

Finally, in Fig.~\ref{fig:fluxlab_smalln}, the energy produced per unit bandwidth and unit time on the rightgoing branch $V_{\om(\Om)}^{\rm out}$ is plotted in the laboratory reference frame.
Quite surprisingly the frequency window of Hawking-like emission is much larger. In fact, even if the comoving frequencies $\ommin$ and $\ommax$ almost coincide, $k$ varies in a wide range, because the comoving-frame group velocity $v_g$ almost vanishes. Consequently, from Eq.~\eqref{eq:boosfrequency}, the laboratory frequency $\Om$ varies in a quite wide window. However, since the total number of created particles must be the same in both reference frames, the production rate per unit bandwidth is much 
smaller. 
Integrating the production rate over frequency and multiplying for $\Delta X_s/V$, with $\Delta X_s\approx 1\,\mbox{mm}$, we obtain a tiny average number ($N\approx 8\times10^{-4}$) of particles produced by each pulse.
In this configuration, fewer particles are produced on the rightgoing mode $V_\om^{\rm out}$ with respect to other modes. In fact, the total number of photons produced on positive-norm modes, obtained by integrating all the curves of Fig.~\ref{fig:fluxall_smalln}, left panel, is $N_{\rm tot}\approx 7\times10^{-2}$.

\subsection{Horizonless configuration}

In this section we investigate a configuration where the velocity of the pulse is so large that no horizon is present, because the two rightmost solutions on the optical branch disappear [see Fig.~\ref{fig:dispsimp} (right panel) or Fig.~\ref{fig:dispersion} (bottom right panel)].
This situation was realized in the experiment of Ref.~\cite{faccioexp}.
A system without any horizon behaves similarly to the case investigated in Sec.~\ref{subsec:abovebelow} for $\om<\ommin$ or $\om>\ommax$, with the exception that there is now no frequency window where eight real-$k$ solutions of the dispersion relation exist in the right region. As a consequence the Hawking-like process cannot occur.
To this purpose, the pulse velocity has been chosen here fast enough that the straight dotted line of the right bottom panel of Fig.~\ref{fig:dispersion} becomes steeper than the tangent to the optical branch at its inflection point.

\begin{figure*}
\centering
\includegraphics{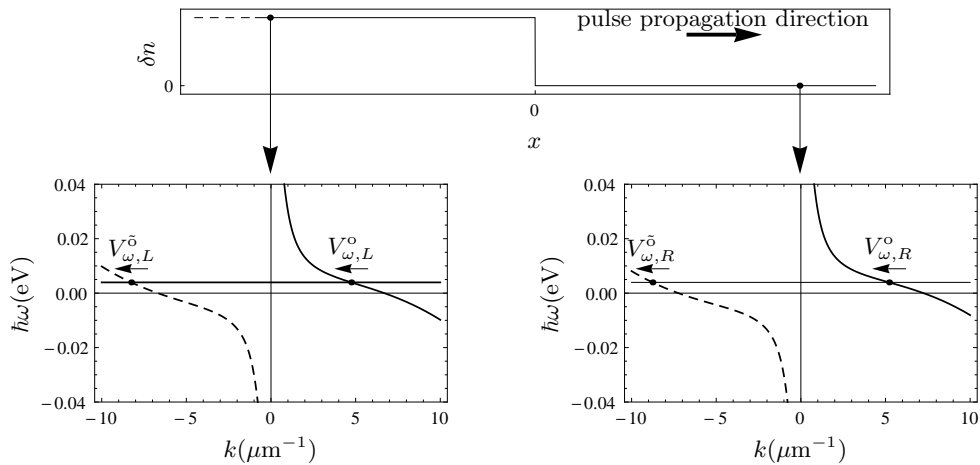}
\caption{{\it Horizonless configuration.} Simplified analysis of the Sellmeier dispersion relation~\eqref{eq:dispersion} in fused silica, as seen from the comoving frame. Only the optical branch is shown.
Positive (negative) laboratory frequency branches are represented by solid (dashed) curves.
The dispersion is plotted on the left ($\delta n=0.001$, left panel) and on the right ($\delta n =0$, right panel) of a perturbation moving rightward with $v=0.69c$ in the laboratory frame (see top panel). For all frequencies there are two real solutions in both the left (left panel) and the right (right panel) regions. Modes are labeled with the notation introduced in Sec.~\ref{sec:settings} and their propagation direction is indicated by arrows.}
\label{fig:dispsimpfaccio}
\end{figure*}

\begin{figure*}
\begin{center}
\includegraphics{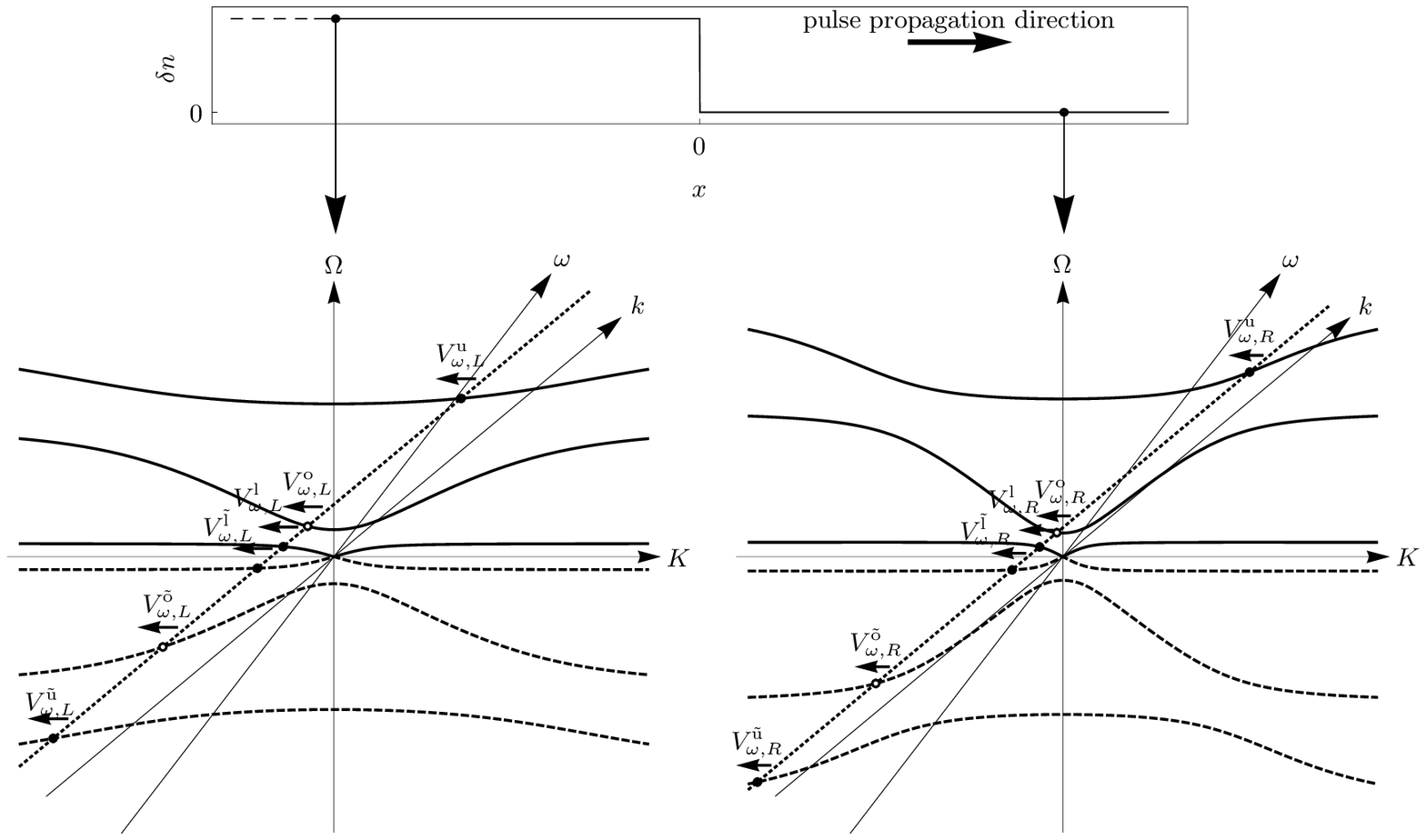}
\end{center}
\caption{{\it Horizonless configuration.} Graphical representation of the Sellmeier dispersion relation, as seen from the laboratory reference frame $(\Om,K)$, for $x<0$ (left panel) and $x>0$ (right panel). The chosen plot range excludes the highest-frequency branch. However, there are no solutions on it (see discussion in Sec.~\ref{subsec:asymptotic}).
The $(\om,k)$ axes of the comoving reference frame are obtained through a boost of velocity $v$.
As sketched in the top panel, the refractive index in the left region is larger than in the right region. This difference in the refractive index is obtained by properly changing the parameters $\beta_i$ and $\Om_i$ appearing in the Lagrangian~\eqref{eq:lagrangian}.
In this plot the values of the velocity $v$ and of the refractive index change $\delta n$ have been arbitrarily chosen for illustrative purposes.
The dispersion relation is graphically solved for a fixed comoving frequency $\om$.
Solutions appear both on the positive-norm positive-$\Om$ branches (solid curves) and on the negative-norm negative-$\Om$ branches (dashed curves).
The open dots denote solutions on the optical branches with positive (o) and negative ($\rm\tilde o$) frequency $\Om$. The arrows indicate the direction of propagation (group velocity in the comoving frame) of the associated modes $V_{\om,L/R}^{\alpha/\tilde\alpha}$.
The dispersion relation has only six real-$k$ solutions in both regions, corresponding to six propagating modes on both sides.}
\label{fig:dispersion_faccio}
\end{figure*}

In analogy with Fig.~\ref{fig:dispsimp}, in Fig.~\ref{fig:dispsimpfaccio} the dispersion relation in the comoving frame is solved only for the optical branch, on both sides of the perturbation, for a pulse with  $\delta n=0.001$ and moving at $v\approx 0.69 c$, which are the parameters of the experiment of Ref.~\cite{faccioexp}. Only two solutions (one with positive and one with negative norm) are present both in the left ($V_{\om,L}^{\rm o}$, $V_{\om,L}^{\rm \tilde o}$) and in the right ($V_{\om,R}^{\rm o}$, $V_{\om,R}^{\rm \tilde o}$) regions for all values of the frequency $\om$.
In particular, the solutions corresponding to AMs $V^{\rm o2}_{\omega,R}$ and $V_{\om,R}$, the latter being the mode responsible for analog Hawking radiation, disappear.
Even if particle production is still possible, due to the mixing of positive- and negative-norm frequency modes, the spectrum will be different and the Hawking-like channel is absent.
For the sake of completeness, in Fig.~\ref{fig:dispersion_faccio}, the full Sellmeier dispersion relation is graphically solved for this configuration, for $x<0$ (left panels) and $x>0$ (right panels).

\begin{figure*}
 \includegraphics{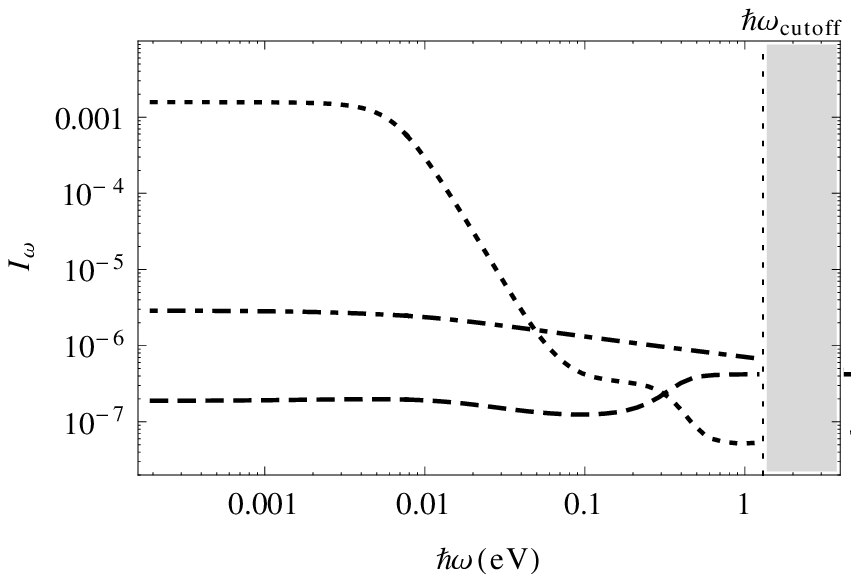}
 \hspace{1.3em}
 \includegraphics{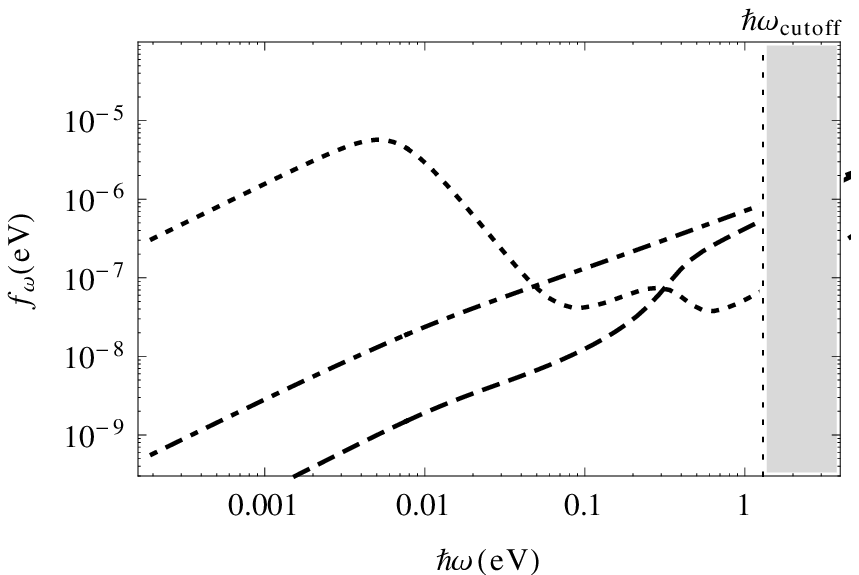}
 \caption{{\it Horizonless configuration.} Occupation numbers $I_\om^{\rm out,l}$ (dashed line), $I_\om^{\rm out,o}$ (dotted line), and $I_\om^{\rm out,u}$ (dot-dashed line) and the corresponding energy fluxes (right panel) $f_\om^{\rm out}$, $f_\om^{\rm out,l}$, $f_\om^{\rm out,o}$, and $f_\om^{\rm out,u}$, for the horizonless configuration of Fig.~\ref{fig:dispersion_faccio}.
The velocity of the pulse and the variation of the refractive index are $v=0.69 c$ and $\epsilon=0.0026$, the parameters of the experiment of Ref.~\cite{faccioexp}, yielding via Eq.~\eqref{eq:betaomleft} a jump in the refractive index of $\delta n=0.001$.}
 \label{fig:flux_faccio}
\end{figure*}

The only effect caused by the absence of the horizon is the disappearance of the flux associated with the outgoing rightgoing mode.
The fluxes of emitted photons (Fig.~\ref{fig:flux_faccio}) are suppressed by the smallness of $\delta n$, as in Sec.~\ref{subsec:smalln}.
The integrated (over frequency) number of created particles ($N\approx0.06$, for a pulse which stably propagates for about 1~mm) does not significantly differ with respect to the horizon configuration with the same small value of $\delta n$.

As a final point it is useful to note that the results of this section aim to reproduce the emission from the leading edge of a propagating pulse. However, it has been pointed out that
in an actual experiment nonlinear effects in pulse propagation make the trailing edge of the pulse much steeper than the leading one~\cite{faccionjp,angus}. Thus, most of the emission should come from the trailing edge. In our formalism this situation corresponds to a steplike profile where the two regions of different refractive index are exchanged with respect to the top panel of Fig.~\ref{fig:dispersion_faccio}. In contrast to horizon configurations, where a black hole would be turned into a white hole, no significant change is expected to appear in the horizonless case, as the topology of modes (six propagating modes in both the right and the left region) remains unchanged. A complete study of configurations corresponding to the trailing edge of a propagating refractive index perturbation (both with and without horizon) will be the subject of a forthcoming publication.

%

\section{Summary and discussion}

\label{sec:conclu}

In this paper we have developed a microscopic theory of the spontaneous quantum vacuum emission generated by a strong light pulse propagating in a Kerr nonlinear optical medium: The effect of the pulse is modeled as a moving refractive index perturbation following in a local and instantaneous way the intensity profile of the strong pulse. For simplicity, we restrict our attention to the case of a single sharp interface separating two homogeneous regions of spatially constant optical properties for which several authors have anticipated the occurrence of the optical analog of a black hole horizon  for suitable values of the pulse speed and the amplitude of the refractive index jump. In contrast to previous work, our theory takes into full account the unavoidable frequency dispersion of the refractive index of the dielectric medium. Describing the leading edge of pulse, we identified the different regimes with or without an analog black hole horizon that can be obtained depending on the pulse parameters. Future work will extend our theory to the emission from the trailing edge of the pulse in both cases, with or without an analog white hole horizon.

Moving to the reference frame comoving with the pulse where the optical properties of the system are time-independent,  the classical eigenmodes of the coupled electromagnetic and matter polarization fields at a given frequency are derived within a Lagrangian formalism and then quantized through canonical quantization. The mixing of positive- and negative-norm modes at the same frequency is responsible for the emission of quantum vacuum radiation, whose rate of production has been computed.

In a configuration with an analog black hole horizon, one can identify a quantum vacuum emission channel showing some similarities with Hawking radiation: This emission channel completely disappears in the absence of a horizon and the corresponding emission temperature has a moderate frequency dependence when observed in the frame comoving with the pulse. 
On the other hand, because of the nontrivial shape of the Sellmeier dispersion relation, this emission channel is only available and active for frequencies in the restricted $\ommin<\om<\ommax$ range where the analog black hole horizon exists, and the emission spectrum dramatically deviates from a thermal law when observed in the laboratory frame.
In addition to this Hawking-like radiation, we have found that quantum vacuum radiation is also emitted on several other channels and outside this frequency window, albeit with a much weaker intensity, a strongly nonthermal spectral distribution, and independently of the presence or absence of the horizon.

As the experiment in Ref.~\cite{faccioexp} was carried out in a parameter range not showing any horizon, it is interesting to conclude the paper by discussing whether it is legitimate to denote the quantum vacuum emission in this regime as ``Hawking radiation.'' This question is all the more relevant given the ongoing debate~\cite{comment,reply} on the interpretation of the experimental results. Even if the results of our calculations are objective, an answer to this question requires a preliminary agreement on the definition of Hawking radiation, which is somehow a matter of personal taste.

If a broad definition is chosen, where the only requirement is the steady production of particles out of the quantum vacuum, there are no difficulties in considering the predicted emission as an example Hawking radiation. The answer is different if we also require the emission to be thermal, which is definitely not the case of the quantum vacuum emission in horizonless configurations or outside the $\ommin<\om<\ommax$ window of a horizon configurations.
Even inside the $\ommin<\om<\ommax$ of a horizon configuration some care has to be paid not to overstretch the gravitational analogy: Even if the emission has an approximately thermal spectrum within this window, the fact that is does not extend down to low frequencies is related to the difficulties of a description of the light propagation in terms of a curved space-time metric that underlies the very concept of the gravitational analogy.

\section{Acknowledgments}
We are grateful to D. Faccio for many discussions and explanations about his experiment~\cite{faccioexp,faccionjp} and for helpful suggestions to improve the readability of the manuscript. We thank R. Balbinot for continuous exchanges and helpful comments about field transformations in a Lagrangian theory, R. Parentani for valuable suggestions about canonical quantization and mode normalization, and S. Liberati and A. Prain for many discussions about the experiment of Ref.~\cite{faccioexp}. We also thank W.~G. Unruh and R. Sch\"utzhold for stimulating discussions during the workshop on Effective Gravity in fluids and superfluids at ICTP, Trieste. This work has been supported by ERC through the QGBE grant.

\appendix

\section{Lagrangian density in the comoving frame}
\label{app:scalars}

In Sec.~\ref{subsec:field}, we applied a Lorentz boost $\Lambda$ to the system described by the Lagrangian of Eq.~\eqref{eq:lagrangian}, to obtain Eq.~\eqref{eq:lagrangianboost}. In doing so, we transformed the space and time coordinates, but we treated both the fields $A$ and $P$ as scalars.\footnote{To maintain the notation compact, the subscript $i$ is omitted in this Appendix.} This is possible since we are dealing with a $1+1$ dimensional system.

To properly proceed, one should first apply the Lorentz transformation $\Lambda$ to the fields $A$ and $P$,
assuming that both $A$ and $P$ oscillate in the $z$ direction. $A$ is the $z$ component of the electromagnetic potential
\begin{equation}\label{eq:amu}
 A^\mu=\begin{pmatrix}
        0\\0\\0\\A
       \end{pmatrix},
\end{equation}
which transforms as
\begin{equation}
 A^\mu_p={\Lambda^\mu}_\nu A^\mu,
\end{equation}
where the subscript $p$ indicates that $A^\mu_p$ is measured in the reference frame comoving with the pulse.
Analogously, $P$ is one of the three components of the magnetization-polarization tensor
\begin{equation}\label{eq:mmunu}
\begin{aligned}
 M^{\mu\nu}&=
 \begin{pmatrix}
  0 & cP_x & cP_y & cP_z\\
  -cP_x & 0 & -M_z & M_y\\
  -cP_y & M_z & 0 & -M_x\\
  -cP_z & -M_y & M_x & 0
 \end{pmatrix}\\
 &=
 \begin{pmatrix}
  0 & 0 & 0 & cP\\
  0 & 0 & 0 & 0\\
  0 & 0 & 0 & 0\\
  -cP & 0 & 0 & 0
 \end{pmatrix},
 \end{aligned}
\end{equation}
which transform as
\begin{equation}
 M_p^{\mu\nu}={\Lambda^{\mu}}_\rho {\Lambda^{\nu}}_\sigma\, M^{\rho\sigma}.
\end{equation}
By doing so, the Lagrangian density can be rewritten in term of the new fields $A^\mu_p$ and $M^{\mu\nu}_p$.

Moreover, the Lagrangian density can always be expressed by using new fields that are functions of $A^\mu_p$ and $M^{\mu\nu}_p$, provided that the transformation to those new fields does not involve any time derivative of $A^\mu_p$ and $M^{\mu\nu}_p$.
Such a transformation generates a canonical transformation on the Hamiltonian variables so that the commutation rules among the new fields and their respective momenta are still canonical.
We chose to apply to $A^\mu_p$ and $M^{\mu\nu}_p$ the linear transformation $\Lambda^{-1}$, not involving any time derivative of the fields. The new fields $A_n^\mu$ and $M_n^{\mu\nu}$ are
\begin{equation}
\begin{aligned}
 &A_{n}^\mu\equiv {(\Lambda^{-1})^\mu}_\nu A_p^\nu= {(\Lambda^{-1})^\mu}_\nu{\Lambda^\nu}_\rho A^\rho= A^\mu,\\
 &M_{n}^{\mu\nu}\equiv {(\Lambda^{-1})^\mu}_\rho {(\Lambda^{-1})^\nu}_\sigma\, M_p^{\rho\sigma}\\
 &\qquad= {(\Lambda^{-1})^\mu}_\rho {(\Lambda^{-1})^\nu}_\sigma{\Lambda^{\rho}}_\alpha {\Lambda^{\sigma}}_\beta M_p^{\alpha\beta}=M^{\mu\nu}.
\end{aligned}
\end{equation}
That is, $A^\mu_n$ and $M^{\mu\nu}_n$
coincide with the old fields $A^\mu$ and $M^{\mu\nu}$, as measured in the laboratory frame. The only nonvanishing components of those fields are $A$ and $P$, as in Eqs.~\eqref{eq:amu} and~\eqref{eq:mmunu}.

\section{Mode normalization}
\label{app:norm}

The scalar product~\eqref{eq:scalar} on two eigenmodes with real frequencies $\om_1$ and $\om_2$ and real momenta $k_{\alpha_1}$ and $k_{\alpha_2}$ is
\begin{multline}\label{eq:scalarfourier}
 \left\langle V_{\om_1}^{\alpha_1},V_{\om_2}^{\alpha_2}\right\rangle
 \!=\!\frac{\ii}{\hbar}\!\int\!\!\dx\,\ee^{-\ii(\om_2-\om_1)t+\ii (k_{\alpha_2}-k_{\alpha_1})x}\, {\bar V}_{\om_1}^{\alpha_1\dagger}\eta{\bar V}_{\om_2}^{\alpha_2}\\
 =2\pi\delta(k_{\alpha_2}-k_{\alpha_1})\,\ee^{-\ii(\om_2-\om_1)t}{\bar V}_{\om_1}^{\alpha_1\dagger}\,\eta\, {\bar V}_{\om_2}^{\alpha_2}.
\end{multline}
This proves that modes with different momenta are orthogonal.
However, for each value of $k$ there are eight solutions $\omega$ of the dispersion relation~\eqref{eq:dispersion}. Thus, there are modes with different frequencies $\om_1\neq\om_2$ which share the same momentum $k_{\alpha_1}=k_{\alpha_2}$.
Yet, those modes are solutions of the eigenvalue problem of Eq.~\eqref{eq:fouriersystem} of the Hermitian matrix ${\cal K}(k_\alpha)$. As a consequence, eigenmodes with different eigenfrequencies $\om_1$ and $\om_2$ are also orthogonal.
Since, at a fixed $k$, $\om$ can take only a finite set of values, it is possible to write
\begin{equation}
  {\bar V}_{\om_1}^{\alpha_1\dagger}\,\eta\,{\bar V}_{\om_2}^{\alpha_2}=
  \delta_{\om_2\om_1}\,{\bar V}_{\om_2}^{\alpha_1\dagger}\,\eta\,{\bar V}_{\om_2}^{\alpha_2}.
\end{equation}
Putting this expression into Eq.~\eqref{eq:scalarfourier}
\begin{equation}\label{eq:deltas}
  \left\langle V_{\om_1}^{\alpha_1},V_{\om_2}^{\alpha_2}\right\rangle
 =2\pi\delta(k_{\alpha_2}-k_{\alpha_1})\,\delta_{\om_2\om_1}\,\frac{\ii}{\hbar}{\bar V}_{\om_2}^{\alpha_1\dagger}\,\eta\,{\bar V}_{\om_2}^{\alpha_2},
\end{equation}
and noting that 
\begin{equation}
 \delta(k_{\alpha_2}-k_{\alpha_1})\,\delta_{\om_2\om_1}=\left|\frac{\dd\om}{\dd k}\right|_{k=k_{\alpha_2}}\delta(\om_2-\om_1)\,\delta_{k_{\alpha_2}k_{\alpha_1}},
\end{equation}
the combination of Dirac and Kroenecker $\delta$'s appearing in Eq.~\eqref{eq:deltas} can be rewritten as
\begin{equation}
 \left|\frac{\dd\om}{\dd k}\right|_{k=k_{\alpha_2}}\delta(\om_2-\om_1)\,\delta_{\alpha_2\alpha_1},
\end{equation}
where, as said, $\alpha$ labels the eight solutions with different momentum, sharing the same comoving frequency $\om$.

In conclusion, we proved that eigenmodes with different $\om$ and $k_\alpha$ are orthogonal:
\begin{equation}\label{eq:resultnorm}
  \left\langle V_{\om_1}^{\alpha_1},V_{\om_2}^{\alpha_2}\right\rangle
 =2\pi\left|\frac{\dd\om}{\dd k}\right|_{k=k_{\alpha_2}}\!\!\!\delta(\om_2-\om_1)\,\delta_{\alpha_2\alpha_1}\,
\frac{\ii}{\hbar} {\bar V}_{\om_2}^{\alpha_2\dagger}\,\eta\,{\bar V}_{\om_2}^{\alpha_2},
\end{equation}
where the last term of this expression is easily computed from Eq.~\eqref{eq:solution}:
\begin{equation}
 \ii\,{\bar V}_{\om}^{\alpha\dagger}\,\eta\,{\bar V}_{\om}^{\alpha}=
 \frac{|C_\om^\alpha|^2}{2\pi}\left[\om+\sum_{i=1}^3\frac{4\pi\beta_i\gamma^2(\om+v k)}{[1-\gamma^2(\om+v k)^2/\Om_i^2]^2} \right],
\end{equation}
where, for the sake of conciseness, the index $\alpha$ is omitted on the right-hand side.
By using the Lorentz transformation~\eqref{eq:boosfrequency},
\begin{multline}\label{eq:VetaV}
 \ii\,{\bar V}_{\om}^{\alpha\dagger}\,\eta\,{\bar V}_{\om}^{\alpha}\\=
 \frac{\gamma|C_\om^\alpha|^2}{2\pi}\left\{-v K+\Om\left[1+\sum_{i=1}^3\frac{4\pi\beta_i}{(1-\Om^2/\Om_i^2)^2} \right]\right\}.\!\!\!
\end{multline}
Note that the scalar product is not positive definite. Thus, we must characterize under what conditions a mode has indeed positive or negative norm.

From the dispersion relation in the glass rest frame~\eqref{eq:sellmeier}, the group velocity $V_g$ in the laboratory frame is given by
\begin{equation}\label{eq:Vg}
 V_g^{-1}
 =\frac{\dd K}{\dd\Om}=\frac{\Om}{K}\frac{\dd K^2}{\dd\Om^2}
 =\frac{\Om}{c^2 K} \left[1 +\sum_{i=1}^3\frac{4\pi\beta_i}{(1-\Om^2/\Om_i^2)^2}\right],
\end{equation}
so that
\begin{equation}\label{eq:resultscalar}
 \ii\,{\bar V}_{\om}^{\alpha\dagger}\,\eta\,{\bar V}_{\om}^{\alpha}=
 \frac{\gamma c^2|C_\om^\alpha|^2}{2\pi}\frac{K}{V_g}
 \left(1-\frac{vV_g}{c^2}\right).
\end{equation}
Since there is no absorption, the group velocity $|V_g|$ must be smaller than the speed of light $c$~\cite{landau8}. Furthermore, $v<c$, so that the term in parentheses in Eq.~\eqref{eq:resultscalar} is always positive. Furthermore, the sign of $V_g$ is positive (negative) if $\Om$ and $K$ have the same (opposite) sign [see Eq.~\eqref{eq:Vg}]. As a consequence, the sign of the above scalar product is always equal to the sign of $\Om$; that is, modes with positive (negative) rest frame frequency $\Om$ have positive (negative) norm.

Finally, the normalization constant $C_\om^\alpha$, appearing in Eq.~\eqref{eq:solution} is fixed by imposing
\begin{equation}
  |\left\langle V_{\om_1}^{\alpha_1},V_{\om_2}^{\alpha_2}\right\rangle|
 =\delta(\om_2-\om_1)\,\delta_{\alpha_2\alpha_1}
\end{equation}
for both positive- and negative-norm modes and by comparing the above condition with Eqs.~\eqref{eq:resultnorm} and~\eqref{eq:resultscalar}:
\begin{equation}
 |C_\om^\alpha|^2
 =
\left|\frac{\gamma c^2}{\hbar}\frac{v_g}{V_g}
 \left(1-\frac{vV_g}{c^2}\right)K\right|^{-1}.
\end{equation}
Using the relativistic composition of velocities
\begin{equation}
 v_g=\frac{V_g-v}{1-v V_g/c^2},
\end{equation}
Eq.~\eqref{eq:resultscalar} simplifies to
\begin{equation}
 |C_\om^\alpha|^2
 =
\left|\frac{\gamma c^2}{\hbar}
\left(1-\frac{v}{V_g}\right)
 K\right|^{-1}.
\end{equation}
Replacing $V_g$ from Eq.~\eqref{eq:Vg},
\begin{equation}
 |C_\om^\alpha|^2
 =
\hbar\left|
c^2\gamma\left(K-\frac{v}{c^2}\Om\right) -
v\sum_{i=1}^3\frac{4\pi\beta_i\gamma\Om_i}{(1-\Om^2/\Om_i^2)^2}
 \right|^{-1},
\end{equation}
and going back to the comoving frame frequency $\om$ and momentum $k_\alpha$,
\begin{equation}
 |C_\om^\alpha|^2
 =
\hbar\left|
c^2 k_\alpha -
v\sum_{i=1}^3\frac{4\pi\beta_i\gamma^2(\om-vk_\alpha)}{[1-\gamma^2(\om-vk_\alpha)^2/\Om_i^2]^2}
 \right|^{-1}.
\end{equation}
%

\section{Emission spectrum for simplified single-pole dispersion relation}
\label{app:thermal}

%
\begin{figure*}
\centering
\includegraphics{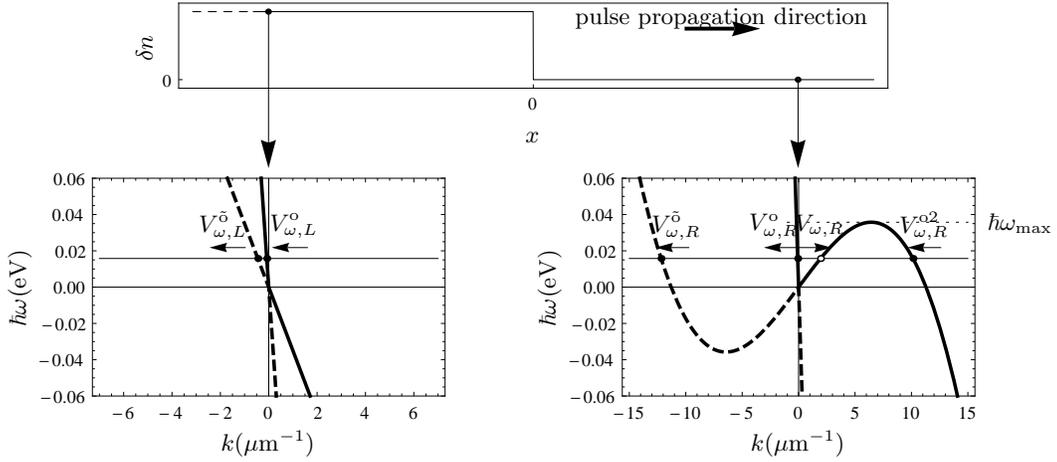}
\caption{{\it Black-hole-like configuration---single-pole dispersion.} Simplified dispersion relation~\eqref{eq:simpdisp} with a single pole as seen from the comoving frame. Only the optical branch is shown.
Positive (negative) laboratory frequency branches are represented by solid (dashed) curves.
The dispersion is plotted on the left ($\delta n=0.12$, left panel, interior of the analog black hole) and on the right ($\delta n =0$, right panel, exterior of the analog black hole) of a perturbation moving rightward with $v=0.83c$ in the laboratory frame (see top panel).
The black horizontal line represents a generic frequency for which there are four real solutions in the right region (right panel), two real solutions in the left one (left panel), and the system shows an analog black hole horizon.
In the right panel, the dashed horizontal line indicates the maximum frequency for which this behavior occurs.
Modes are labeled with the notation introduced in Sec.~\ref{sec:settings} and their propagation direction is indicated by arrows.}
\label{fig:dispsimpsimp}
\end{figure*}
\begin{figure*}
 \includegraphics{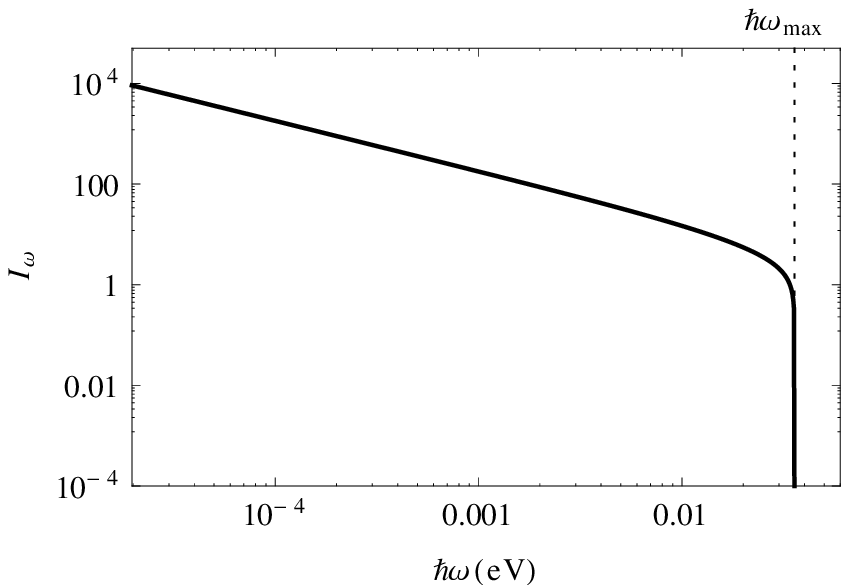}
 \hspace{1.3em}
 \includegraphics{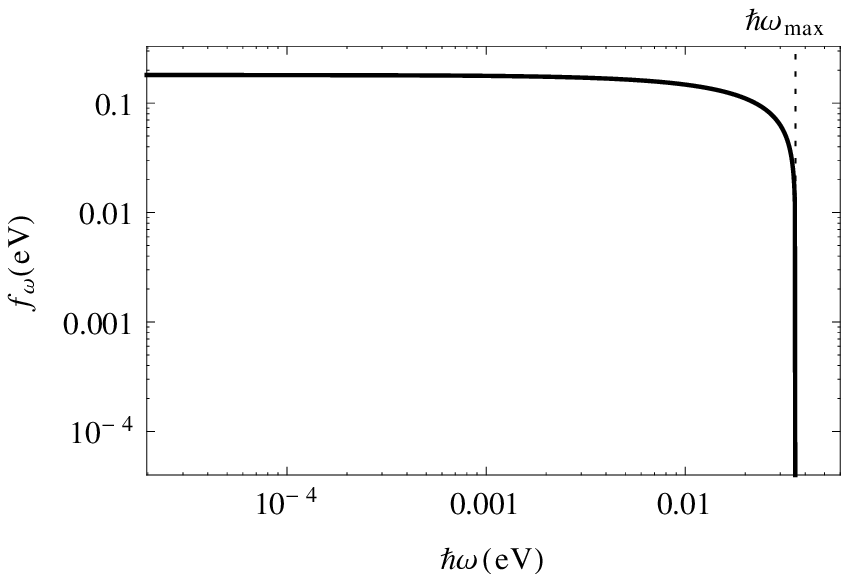}
 \caption{{\it Black-hole-like configuration---single-pole dispersion.} Flux of rightgoing particles emitted in the external region of the analog black hole (left panel) $I_\om^{\rm out}$ and the corresponding energy flux (right panel) $f_\om^{\rm out}$, as seen from the comoving reference frame, for a black hole horizon configuration as sketched in Fig.~\ref{fig:dispsimpsimp} with a simplified single-pole dispersion relation of the form of Eq.~\eqref{eq:simpdisp}. Parameters $v=0.83c$ and $\epsilon=0.3$, yielding via Eq.~\eqref{eq:betaomleft} a refractive index jump of $\delta n=0.12$ for optical frequencies.}
 \label{fig:fluxthermal}
\end{figure*}

To further assess the importance of using a realistic form of the dispersion relation to study analog Hawking emission, in this appendix we apply our method to calculate the emission spectrum for the case of the simplified dispersion relation with a single pole
\begin{equation}\label{eq:simpdisp}
  c^2 K^2 = \Om^2\left[1 +\frac{4\pi\beta}{1-\Om^2/\Om_0^2}\right],
\end{equation}
In order for the resulting standard polariton dispersion~\cite{hopfield} to approximate the full Sellmeier one~\eqref{eq:sellmeier} in the optical frequency range, we choose $\beta=\beta_2$ and $\Om_0=\Om_2$. Since the analog geometry is now well defined in the low-frequency limit, black hole horizon configurations are easily identified, as sketched in Fig.~\ref{fig:dispsimpsimp}. In general, for a fixed comoving frequency $\om$ this dispersion relation Eq.~\eqref{eq:simpdisp} has four solutions on both sides of the discontinuity. When an analog black hole horizon is present, two of these solutions are real and two are complex conjugate in the region corresponding to the interior of the black hole, while all four solutions are real in the region representing the exterior of the black hole.

Below a certain cutoff frequency $\ommax$, a Hawking-like channel is present. In Fig.~\ref{fig:fluxthermal} we plot the flux $I_\om^{\rm out}$ of rightgoing particles, emitted in the external (right) region of the analog black hole, and the corresponding energy flux $f_\om^{\rm out}$ on this channel. The flux of particles diverges as $1/\om$ for $\om\to0$, while the flux of energy is constant as predicted for a thermal emission. This agreement with the original Hawking's prediction in this simplified situation could be expected as the analogy with black hole physics is exact in the low-frequency/long-wavelength limit.
Consequently, in materials where the low-frequency pole in the infrared domain is not present and the only poles are in the ultraviolet (for instance in diamond~\cite{refractiveindex}), it should be possible to experimentally assess the thermal nature of Hawking radiation.

Comparison of these results with the predictions for the full Sellmeier dispersion discussed in Sec.~\ref{sec:emission} points out the crucial role of the low-frequency pole in the infrared domain, that prevents the identification of an analog geometry in the optical frequency range. 
This is the physical reason for the strongly modified spectrum with respect to the simpler forms of subluminal dispersion relations considered, for example, in~\cite{Scott_thesis,lr}. Of course, also in this case spontaneous emission is still present in horizonless configurations but with a strongly nonthermal spectrum, as found in an analog model based on Bose--Einstein condensates~\cite{granada_proc}.

\bibliography{kinetics}

\begin{thebibliography}{10}%
\makeatletter
\providecommand \@ifxundefined [1]{%
 \ifx #1\undefined \expandafter \@firstoftwo
 \else \expandafter \@secondoftwo
\fi
}%
\providecommand \@ifnum [1]{%
 \ifnum #1\expandafter \@firstoftwo
 \else \expandafter \@secondoftwo
\fi
}%
\providecommand \enquote [1]{``#1''}%
\providecommand \bibnamefont  [1]{#1}%
\providecommand \bibfnamefont [1]{#1}%
\providecommand \citenamefont [1]{#1}%
\providecommand\href[0]{\@sanitize\@href}%
\providecommand\@href[1]{\endgroup\@@startlink{#1}\endgroup\@@href}%
\providecommand\@@href[1]{#1\@@endlink}%
\providecommand \@sanitize [0]{\begingroup\catcode`\&12\catcode`\#12\relax}%
\@ifxundefined \pdfoutput {\@firstoftwo}{%
 \@ifnum{\z@=\pdfoutput}{\@firstoftwo}{\@secondoftwo}%
}{%
 \providecommand\@@startlink[1]{\leavevmode\special{html:<a href="#1">}}%
 \providecommand\@@endlink[0]{\special{html:</a>}}%
}{%
 \providecommand\@@startlink[1]{%
  \leavevmode
  \pdfstartlink
   attr{/Border[0 0 1 ]/H/I/C[0 1 1]}%
   user{/Subtype/Link/A<</Type/Action/S/URI/URI(#1)>>}%
  \relax
 }%
 \providecommand\@@endlink[0]{\pdfendlink}%
}%
\providecommand \url  [0]{\begingroup\@sanitize \@url }%
\providecommand \@url [1]{\endgroup\@href {#1}{\urlprefix}}%
\providecommand \urlprefix [0]{URL }%
\providecommand \Eprint[0]{\href }%
\@ifxundefined \urlstyle {%
  \providecommand \doi [1]{doi:\discretionary{}{}{}#1}%
}{%
  \providecommand \doi [0]{doi:\discretionary{}{}{}\begingroup
  \urlstyle{rm}\Url }%
}%
\providecommand \doibase [0]{http://dx.doi.org/}%
\providecommand \Doi[1]{\href{\doibase#1}}%
\providecommand \bibAnnote [3]{%
  \BibitemShut{#1}%
  \begin{quotation}\noindent
    \textsc{Key:}\ #2\\\textsc{Annotation:}\ #3%
  \end{quotation}%
}%
\providecommand \bibAnnoteFile [2]{%
  \IfFileExists{#2}{\bibAnnote {#1} {#2} {\input{#2}}}{}%
}%
\providecommand \typeout [0]{\immediate \write \m@ne }%
\providecommand \selectlanguage [0]{\@gobble}%
\providecommand \bibinfo [0]{\@secondoftwo}%
\providecommand \bibfield [0]{\@secondoftwo}%
\providecommand \translation [1]{[#1]}%
\providecommand \BibitemOpen[0]{}%
\providecommand \bibitemStop [0]{}%
\providecommand \bibitemNoStop [0]{.\EOS\space}%
\providecommand \EOS [0]{\spacefactor3000\relax}%
\providecommand \BibitemShut [1]{\csname bibitem#1\endcsname}%
\bibitem{hawkingnat}%
  \BibitemOpen
  \bibfield{author}{%
  \bibinfo {author} {\bibfnamefont{S.~W.}\ \bibnamefont{{Hawking}}},\ }%
  \bibfield{journal}{%
  \Doi{10.1038/248030a0}{\bibinfo {journal} {Nature (London)}}\ }%
  \textbf{\bibinfo {volume} {248}},\ \bibinfo {pages} {30} (\bibinfo {year}
  {1974})%
  \bibAnnoteFile{NoStop}{hawkingnat}%
\bibitem{hawking75}%
  \BibitemOpen
  \bibfield{author}{%
  \bibinfo {author} {\bibfnamefont{S.~W.}\ \bibnamefont{Hawking}},\ }%
  \bibfield{journal}{%
  \Doi{10.1007/BF02345020}{\bibinfo {journal} {Commun. Math. Phys.}}\ }%
  \textbf{\bibinfo {volume} {43}},\ \bibinfo {pages} {199} (\bibinfo {year}
  {1975})%
  \bibAnnoteFile{NoStop}{hawking75}%
\bibitem{unruh}%
  \BibitemOpen
  \bibfield{author}{%
  \bibinfo {author} {\bibfnamefont{W.~G.}\ \bibnamefont{Unruh}},\ }%
  \bibfield{journal}{%
  \Doi{10.1103/PhysRevLett.46.1351}{\bibinfo {journal} {Phys. Rev. Lett.}}\ }%
  \textbf{\bibinfo {volume} {46}},\ \bibinfo {pages} {1351} (\bibinfo {year}
  {1981})%
  \bibAnnoteFile{NoStop}{unruh}%
\bibitem{lr}%
  \BibitemOpen
  \bibfield{author}{%
  \bibinfo {author} {\bibfnamefont{C.}~\bibnamefont{Barcelo}}, \bibinfo
  {author} {\bibfnamefont{S.}~\bibnamefont{Liberati}},\ and\ \bibinfo {author}
  {\bibfnamefont{M.}~\bibnamefont{Visser}},\ }%
  \bibfield{journal}{%
  \href{http://www.livingreviews.org/lrr-2011-3}{\bibinfo {journal} {Living Rev. Rel.}\ }}%
  \textbf{\bibinfo {volume} {14}},\ \bibinfo {pages} {3} (\bibinfo {year}
  {2011})%
  \bibAnnoteFile{NoStop}{lr}%
\bibitem{ulf_science}%
  \BibitemOpen
  \bibfield{author}{%
  \bibinfo {author} {\bibfnamefont{T.~G.}\ \bibnamefont{{Philbin}}}, \bibinfo
  {author} {\bibfnamefont{C.}~\bibnamefont{{Kuklewicz}}}, \bibinfo {author}
  {\bibfnamefont{S.}~\bibnamefont{{Robertson}}}, \bibinfo {author}
  {\bibfnamefont{S.}~\bibnamefont{{Hill}}}, \bibinfo {author}
  {\bibfnamefont{F.}~\bibnamefont{{K{\"o}nig}}},\ and\ \bibinfo {author}
  {\bibfnamefont{U.}~\bibnamefont{{Leonhardt}}},\ }%
  \bibfield{journal}{%
  \Doi{10.1126/science.1153625}{\bibinfo {journal} {Science}}\ }%
  \textbf{\bibinfo {volume} {319}},\ \bibinfo {pages} {1367} (\bibinfo {year}
  {2008})%
  \bibAnnoteFile{NoStop}{ulf_science}%
\bibitem{transfoptics}%
  \BibitemOpen
  \bibfield{author}{%
  \bibinfo {author} {\bibfnamefont{U.}~\bibnamefont{Leonhardt}}\ and\ \bibinfo
  {author} {\bibfnamefont{T.~G.}\ \bibnamefont{Philbin}},\ }%
  \bibinfo {series} {Progr. Opt.}\ \textbf{\bibinfo {volume} {53}},\ \bibinfo
  {pages} {69} (\bibinfo {year} {2009})%
  \bibAnnoteFile{NoStop}{transfoptics}%
\bibitem{dario}%
  \BibitemOpen
  \bibfield{author}{%
  \bibinfo {author} {\bibfnamefont{D.}~\bibnamefont{{Gerace}}}\ and\ \bibinfo
  {author} {\bibfnamefont{I.}~\bibnamefont{{Carusotto}}},\ }%
  \bibfield{journal}{%
  \Doi{10.1103/PhysRevB.86.144505}{\bibinfo {journal} {\prb}}\ }%
  \textbf{\bibinfo {volume} {86}},\ \bibinfo {eid} {144505} (\bibinfo {year}
  {2012})%
  \bibAnnoteFile{NoStop}{dario}%
\bibitem{fleurovbarad}%
  \BibitemOpen
  \bibfield{author}{%
  \bibinfo {author} {\bibfnamefont{M.}~\bibnamefont{{Elazar}}}, \bibinfo
  {author} {\bibfnamefont{V.}~\bibnamefont{{Fleurov}}},\ and\ \bibinfo {author}
  {\bibfnamefont{S.}~\bibnamefont{{Bar-Ad}}},\ }%
  \bibfield{journal}{%
  \Doi{10.1103/PhysRevA.86.063821}{\bibinfo {journal} {\pra}}\ }%
  \textbf{\bibinfo {volume} {86}},\ \bibinfo {eid} {063821} (\bibinfo {year}
  {2012})%
  \bibAnnoteFile{NoStop}{fleurovbarad}%
\bibitem{faccioexp}%
  \BibitemOpen
  \bibfield{author}{%
  \bibinfo {author} {\bibfnamefont{F.}~\bibnamefont{Belgiorno}}, \bibinfo
  {author} {\bibfnamefont{S.~L.}\ \bibnamefont{Cacciatori}}, \bibinfo {author}
  {\bibfnamefont{M.}~\bibnamefont{Clerici}}, \bibinfo {author}
  {\bibfnamefont{V.}~\bibnamefont{Gorini}}, \bibinfo {author}
  {\bibfnamefont{G.}~\bibnamefont{Ortenzi}}, \bibinfo {author}
  {\bibfnamefont{L.}~\bibnamefont{Rizzi}}, \bibinfo {author}
  {\bibfnamefont{E.}~\bibnamefont{Rubino}}, \bibinfo {author}
  {\bibfnamefont{V.~G.}\ \bibnamefont{Sala}},\ and\ \bibinfo {author}
  {\bibfnamefont{D.}~\bibnamefont{Faccio}},\ }%
  \bibfield{journal}{%
  \Doi{10.1103/PhysRevLett.105.203901}{\bibinfo {journal} {Phys. Rev. Lett.}}\
  }%
  \textbf{\bibinfo {volume} {105}},\ \bibinfo {pages} {203901} (\bibinfo {year}
  {2010})%
  \bibAnnoteFile{NoStop}{faccioexp}%
\bibitem{faccionjp}%
  \BibitemOpen
  \bibfield{author}{%
  \bibinfo {author} {\bibfnamefont{E.}~\bibnamefont{{Rubino}}}, \bibinfo
  {author} {\bibfnamefont{F.}~\bibnamefont{{Belgiorno}}}, \bibinfo {author}
  {\bibfnamefont{S.~L.}\ \bibnamefont{{Cacciatori}}}, \bibinfo {author}
  {\bibfnamefont{M.}~\bibnamefont{{Clerici}}}, \bibinfo {author}
  {\bibfnamefont{V.}~\bibnamefont{{Gorini}}}, \bibinfo {author}
  {\bibfnamefont{G.}~\bibnamefont{{Ortenzi}}}, \bibinfo {author}
  {\bibfnamefont{L.}~\bibnamefont{{Rizzi}}}, \bibinfo {author}
  {\bibfnamefont{V.~G.}\ \bibnamefont{{Sala}}}, \bibinfo {author}
  {\bibfnamefont{M.}~\bibnamefont{{Kolesik}}},\ and\ \bibinfo {author}
  {\bibfnamefont{D.}~\bibnamefont{{Faccio}}},\ }%
  \bibfield{journal}{%
  \Doi{10.1088/1367-2630/13/8/085005}{\bibinfo {journal} {New J. Phys.}}\ }%
  \textbf{\bibinfo {volume} {13}},\ \bibinfo {pages} {085005} (\bibinfo {year}
  {2011})%
  \bibAnnoteFile{NoStop}{faccionjp}%
\bibitem{durninprl}%
  \BibitemOpen
  \bibfield{author}{%
  \bibinfo {author} {\bibfnamefont{J.}~\bibnamefont{{Durnin}}}, \bibinfo
  {author} {\bibfnamefont{J.~J.}\ \bibnamefont{{Miceli}}, \bibfnamefont{Jr.}},\
  and\ \bibinfo {author} {\bibfnamefont{J.~H.}\ \bibnamefont{{Eberly}}},\ }%
  \bibfield{journal}{%
  \Doi{10.1103/PhysRevLett.58.1499}{\bibinfo {journal} {Phys. Rev. Lett.}}\ }%
  \textbf{\bibinfo {volume} {58}},\ \bibinfo {pages} {1499} (\bibinfo {year}
  {1987})%
  \bibAnnoteFile{NoStop}{durninprl}%
\bibitem{gori}%
  \BibitemOpen
  \bibfield{author}{%
  \bibinfo {author} {\bibfnamefont{F.}~\bibnamefont{{Gori}}}, \bibinfo {author}
  {\bibfnamefont{G.}~\bibnamefont{{Guattari}}},\ and\ \bibinfo {author}
  {\bibfnamefont{C.}~\bibnamefont{{Padovani}}},\ }%
  \bibfield{journal}{%
  \Doi{10.1016/0030-4018(87)90276-8}{\bibinfo {journal} {Opt. Commun.}}\ }%
  \textbf{\bibinfo {volume} {64}},\ \bibinfo {pages} {491} (\bibinfo {year}
  {1987})%
  \bibAnnoteFile{NoStop}{gori}%
\bibitem{mcdonald}%
  \BibitemOpen
  \bibfield{author}{%
  \bibinfo {author} {\bibfnamefont{K.~T.}\ \bibnamefont{{McDonald}}}\ }%
  \Eprint{http://arxiv.org/abs/physics/0006046}{arXiv:physics/0006046}%
  \bibAnnoteFile{NoStop}{mcdonald}%
\bibitem{comment}%
  \BibitemOpen
  \bibfield{author}{%
  \bibinfo {author} {\bibfnamefont{R.}~\bibnamefont{Sch\"utzhold}}\ and\
  \bibinfo {author} {\bibfnamefont{W.~G.}\ \bibnamefont{Unruh}},\ }%
  \bibfield{journal}{%
  \Doi{10.1103/PhysRevLett.107.149401}{\bibinfo {journal} {Phys. Rev. Lett.}}\
  }%
  \textbf{\bibinfo {volume} {107}},\ \bibinfo {pages} {149401} (\bibinfo {year}
  {2011})%
  \bibAnnoteFile{NoStop}{comment}%
\bibitem{reply}%
  \BibitemOpen
  \bibfield{author}{%
  \bibinfo {author} {\bibfnamefont{F.}~\bibnamefont{Belgiorno}}, \bibinfo
  {author} {\bibfnamefont{S.~L.}\ \bibnamefont{Cacciatori}}, \bibinfo {author}
  {\bibfnamefont{M.}~\bibnamefont{Clerici}}, \bibinfo {author}
  {\bibfnamefont{V.}~\bibnamefont{Gorini}}, \bibinfo {author}
  {\bibfnamefont{G.}~\bibnamefont{Ortenzi}}, \bibinfo {author}
  {\bibfnamefont{L.}~\bibnamefont{Rizzi}}, \bibinfo {author}
  {\bibfnamefont{E.}~\bibnamefont{Rubino}}, \bibinfo {author}
  {\bibfnamefont{V.~G.}\ \bibnamefont{Sala}},\ and\ \bibinfo {author}
  {\bibfnamefont{D.}~\bibnamefont{Faccio}},\ }%
  \bibfield{journal}{%
  \Doi{10.1103/PhysRevLett.107.149402}{\bibinfo {journal} {Phys. Rev. Lett.}}\
  }%
  \textbf{\bibinfo {volume} {107}},\ \bibinfo {pages} {149402} (\bibinfo {year}
  {2011})%
  \bibAnnoteFile{NoStop}{reply}%
\bibitem{angus}%
  \BibitemOpen
  \bibfield{author}{%
  \bibinfo {author} {\bibfnamefont{S.}~\bibnamefont{{Liberati}}}, \bibinfo
  {author} {\bibfnamefont{A.}~\bibnamefont{{Prain}}},\ and\ \bibinfo {author}
  {\bibfnamefont{M.}~\bibnamefont{{Visser}}},\ }%
  \bibfield{journal}{%
  \Doi{10.1103/PhysRevD.85.084014}{\bibinfo {journal} {\prd}}\ }%
  \textbf{\bibinfo {volume} {85}},\ \bibinfo {eid} {084014} (\bibinfo {year}
  {2012})%
  \bibAnnoteFile{NoStop}{angus}%
\bibitem{unruhuniverse}%
  \BibitemOpen
  \bibfield{author}{%
  \bibinfo {author} {\bibfnamefont{W.~G.}\ \bibnamefont{{Unruh}}}\ and\
  \bibinfo {author} {\bibfnamefont{R.}~\bibnamefont{{Sch{\"u}tzhold}}},\ }%
  \bibfield{journal}{%
  \Doi{10.1103/PhysRevD.86.064006}{\bibinfo {journal} {\prd}}\ }%
  \textbf{\bibinfo {volume} {86}},\ \bibinfo {eid} {064006} (\bibinfo {year}
  {2012})%
  \bibAnnoteFile{NoStop}{unruhuniverse}%
\bibitem{Scott_thesis}%
  \BibitemOpen
  \bibfield{author}{%
  \bibinfo {author} {\bibfnamefont{S.~J.}\ \bibnamefont{{Robertson}}},\ }%
  \enquote{\bibinfo {title} {{Ph.D. thesis, School of Physics and Astronomy,
  University of St Andrews, 2011}},}\
  \Eprint{http://arxiv.org/abs/1106.1805}{arXiv:1106.1805 [gr-qc]}%
  \bibAnnoteFile{NoStop}{Scott_thesis}%
\bibitem{robertson}%
  \BibitemOpen
  \bibfield{author}{%
  \bibinfo {author} {\bibfnamefont{U.}~\bibnamefont{{Leonhardt}}}\ and\
  \bibinfo {author} {\bibfnamefont{S.}~\bibnamefont{{Robertson}}},\ }%
  \bibfield{journal}{%
  \Doi{10.1088/1367-2630/14/5/053003}{\bibinfo {journal} {New J. Phys.}}\ }%
  \textbf{\bibinfo {volume} {14}},\ \bibinfo {pages} {053003} (\bibinfo {year}
  {2012})%
  \bibAnnoteFile{NoStop}{robertson}%
\bibitem{kinetics}%
  \BibitemOpen
  \bibfield{author}{%
  \bibinfo {author} {\bibfnamefont{S.}~\bibnamefont{{Finazzi}}}\ and\ \bibinfo
  {author} {\bibfnamefont{I.}~\bibnamefont{{Carusotto}}},\ }%
  \bibfield{journal}{%
  \Doi{10.1140/epjp/i2012-12078-x}{\bibinfo {journal} {Eur. Phys. J. Plus}}\ }%
  \textbf{\bibinfo {volume} {127}},\ \bibinfo {pages} {78} (\bibinfo {year}
  {2012})%
  \bibAnnoteFile{NoStop}{kinetics}%
\bibitem{butchercotter}%
  \BibitemOpen
  \bibfield{author}{%
  \bibinfo {author} {\bibfnamefont{P.}~\bibnamefont{Butcher}}\ and\ \bibinfo
  {author} {\bibfnamefont{D.}~\bibnamefont{Cotter}},\ }%
  \emph{\bibinfo {title} {The Elements of Nonlinear Optics}}\ (\bibinfo
  {publisher} {Cambridge University Press},\ \bibinfo {address} {Cambridge},\
  \bibinfo {year} {1991})%
  \bibAnnoteFile{NoStop}{butchercotter}%
\bibitem{hopfield}%
  \BibitemOpen
  \bibfield{author}{%
  \bibinfo {author} {\bibfnamefont{J.~J.}\ \bibnamefont{{Hopfield}}},\ }%
  \bibfield{journal}{%
  \Doi{10.1103/PhysRev.112.1555}{\bibinfo {journal} {Phys. Rev.}}\ }%
  \textbf{\bibinfo {volume} {112}},\ \bibinfo {pages} {1555} (\bibinfo {year}
  {1958})%
  \bibAnnoteFile{NoStop}{hopfield}%
\bibitem{iacopothree}%
  \BibitemOpen
  \bibfield{author}{%
  \bibinfo {author} {\bibfnamefont{I.}~\bibnamefont{{Carusotto}}}, \bibinfo
  {author} {\bibfnamefont{M.}~\bibnamefont{{Antezza}}}, \bibinfo {author}
  {\bibfnamefont{F.}~\bibnamefont{{Bariani}}}, \bibinfo {author}
  {\bibfnamefont{S.}~\bibnamefont{{de Liberato}}},\ and\ \bibinfo {author}
  {\bibfnamefont{C.}~\bibnamefont{{Ciuti}}},\ }%
  \bibfield{journal}{%
  \Doi{10.1103/PhysRevA.77.063621}{\bibinfo {journal} {\pra}}\ }%
  \textbf{\bibinfo {volume} {77}},\ \bibinfo {eid} {063621} (\bibinfo {year}
  {2008})%
  \bibAnnoteFile{NoStop}{iacopothree}%
\bibitem{subband}%
  \BibitemOpen
  \bibfield{author}{%
  \bibinfo {author} {\bibfnamefont{C.}~\bibnamefont{{Ciuti}}}, \bibinfo
  {author} {\bibfnamefont{G.}~\bibnamefont{{Bastard}}},\ and\ \bibinfo {author}
  {\bibfnamefont{I.}~\bibnamefont{{Carusotto}}},\ }%
  \bibfield{journal}{%
  \Doi{10.1103/PhysRevB.72.115303}{\bibinfo {journal} {\prb}}\ }%
  \textbf{\bibinfo {volume} {72}},\ \bibinfo {eid} {115303} (\bibinfo {year}
  {2005})%
  \bibAnnoteFile{NoStop}{subband}%
\bibitem{iacopo_PRA_backreaction}%
  \BibitemOpen
  \bibfield{author}{%
  \bibinfo {author} {\bibfnamefont{I.}~\bibnamefont{{Carusotto}}}, \bibinfo
  {author} {\bibfnamefont{S.}~\bibnamefont{{de Liberato}}}, \bibinfo {author}
  {\bibfnamefont{D.}~\bibnamefont{{Gerace}}},\ and\ \bibinfo {author}
  {\bibfnamefont{C.}~\bibnamefont{{Ciuti}}},\ }%
  \bibfield{journal}{%
  \Doi{10.1103/PhysRevA.85.023805}{\bibinfo {journal} {\pra}}\ }%
  \textbf{\bibinfo {volume} {85}},\ \bibinfo {eid} {023805} (\bibinfo {year}
  {2012})%
  \bibAnnoteFile{NoStop}{iacopo_PRA_backreaction}%
\bibitem{sellmeier}%
  \BibitemOpen
  \bibfield{author}{%
  \bibinfo {author} {\bibfnamefont{W.}~\bibnamefont{{Sellmeier}}},\ }%
  \bibfield{journal}{%
  \Doi{10.1002/andp.18712190612}{\bibinfo {journal} {Ann. Phys.}}\ }%
  \textbf{\bibinfo {volume} {219}},\ \bibinfo {pages} {272} (\bibinfo {year}
  {1871})%
  \bibAnnoteFile{NoStop}{sellmeier}%
\bibitem{refractiveindex}%
  \BibitemOpen
  \enquote{\bibinfo {title} {Refractive index database},}\
  \url{http://refractiveindex.info}%
  \bibAnnoteFile{NoStop}{refractiveindex}%
\bibitem{2dplots}%
  \BibitemOpen
  \bibfield{author}{%
  \bibinfo {author} {\bibfnamefont{S.}~\bibnamefont{Finazzi}}\ and\ \bibinfo
  {author} {\bibfnamefont{R.}~\bibnamefont{Parentani}},\ }%
  \bibfield{journal}{%
  \Doi{10.1103/PhysRevD.85.124027}{\bibinfo {journal} {Phys. Rev. D}}\ }%
  \textbf{\bibinfo {volume} {85}},\ \bibinfo {pages} {124027} (\bibinfo {year}
  {2012})%
  \bibAnnoteFile{NoStop}{2dplots}%
\bibitem{MacherRP1}%
  \BibitemOpen
  \bibfield{author}{%
  \bibinfo {author} {\bibfnamefont{J.}~\bibnamefont{Macher}}\ and\ \bibinfo
  {author} {\bibfnamefont{R.}~\bibnamefont{Parentani}},\ }%
  \bibfield{journal}{%
  \Doi{10.1103/PhysRevD.79.124008}{\bibinfo {journal} {Phys. Rev. D}}\ }%
  \textbf{\bibinfo {volume} {79}},\ \bibinfo {pages} {124008} (\bibinfo {year}
  {2009})%
  \bibAnnoteFile{NoStop}{MacherRP1}%
\bibitem{Recati2009}%
  \BibitemOpen
  \bibfield{author}{%
  \bibinfo {author} {\bibfnamefont{A.}~\bibnamefont{Recati}}, \bibinfo {author}
  {\bibfnamefont{N.}~\bibnamefont{Pavloff}},\ and\ \bibinfo {author}
  {\bibfnamefont{I.}~\bibnamefont{Carusotto}},\ }%
  \bibfield{journal}{%
  \Doi{10.1103/PhysRevA.80.043603}{\bibinfo {journal} {Phys. Rev. A}}\ }%
  \textbf{\bibinfo {volume} {80}},\ \bibinfo {pages} {043603} (\bibinfo {year}
  {2009})%
  \bibAnnoteFile{NoStop}{Recati2009}%
\bibitem{MacherBEC}%
  \BibitemOpen
  \bibfield{author}{%
  \bibinfo {author} {\bibfnamefont{J.}~\bibnamefont{{Macher}}}\ and\ \bibinfo
  {author} {\bibfnamefont{R.}~\bibnamefont{{Parentani}}},\ }%
  \bibfield{journal}{%
  \Doi{10.1103/PhysRevA.80.043601}{\bibinfo {journal} {Phys. Rev. A}}\ }%
  \textbf{\bibinfo {volume} {80}},\ \bibinfo {pages} {043601} (\bibinfo {month}
  {Oct.}\ \bibinfo {year} {2009})%
  \bibAnnoteFile{NoStop}{MacherBEC}%
\bibitem{landau8}%
  \BibitemOpen
  \bibfield{author}{%
  \bibinfo {author} {\bibfnamefont{L.}~\bibnamefont{Landau}}, \bibinfo {author}
  {\bibfnamefont{E.}~\bibnamefont{Lifshitz}},\ and\ \bibinfo {author}
  {\bibfnamefont{L.}~\bibnamefont{Pitaevski{\u\i}}},\ }%
  \emph{\bibinfo {title} {Electrodynamics of continuous media}},\ Course of
  theoretical physics\ (\bibinfo {publisher} {Butterworth-Heinemann},\ \bibinfo
  {address} {Oxford, UK},\ \bibinfo {year} {1984})%
  \bibAnnoteFile{NoStop}{landau8}%
\bibitem{granada_proc}%
  \BibitemOpen
  \bibfield{author}{%
  \bibinfo {author} {\bibfnamefont{S.}~\bibnamefont{{Finazzi}}}\ and\ \bibinfo
  {author} {\bibfnamefont{R.}~\bibnamefont{{Parentani}}},\ }%
  \bibfield{journal}{%
  \Doi{10.1088/1742-6596/314/1/012030}{\bibinfo {journal} {J. Phys.: Conf.
  Ser.}}\ }%
  \textbf{\bibinfo {volume} {314}},\ \bibinfo {pages} {012030} (\bibinfo {year}
  {2011})%
  \bibAnnoteFile{NoStop}{granada_proc}%
\end{thebibliography}%

\end{document}